\documentclass[twocolumn]{aastex62}

\usepackage{natbib}

\usepackage{graphicx}
\usepackage{epstopdf}
\usepackage{amsmath}
\usepackage{amssymb}
\usepackage{bm}
\usepackage{color}
\usepackage{multirow}
\usepackage{rotating}

\def\gs{\mathrel{\raise0.35ex\hbox{$\scriptstyle >$}\kern-0.6em
\lower0.40ex\hbox{{$\scriptstyle \sim$}}}}
\def\ls{\mathrel{\raise0.35ex\hbox{$\scriptstyle <$}\kern-0.6em
\lower0.40ex\hbox{{$\scriptstyle \sim$}}}}

\newcommand{\Hb}{H$\beta$}

\newcommand{\heii}{He\,{\sc ii}}
\newcommand{\hei}{He\,{\sc i}}
\newcommand{\feii}{Fe\,{\sc ii}}

\newcommand{\oiii}{[O\,{\sc iii}]}

\newcommand{\bg}{BG92}
\newcommand{\kov}{K10}
\newcommand{\veron}{VC04}

\newcommand{\logfmeansigma}{$\log_{10}(f_{{\rm mean},{\sigma}})$}
\newcommand{\logfrmssigma}{$\log_{10}(f_{{\rm rms},{\sigma}})$}
\newcommand{\logfmeanfwhm}{$\log_{10}(f_{{\rm mean},{\rm FWHM}})$}

\newcommand{\meanlogfrmssigma}{$\log_{10}(\bar{f}_{{\rm rms},{\sigma}})$}

\newcommand{\iisa}{PG~1310$-$108}
\newcommand{\zwia}{Zw~229$-$015}
\newcommand{\ngca}{NGC~4593}
\newcommand{\mrka}{Mrk~50}
\newcommand{\mrkb}{Mrk~141}
\newcommand{\mrkc}{Mrk~279}
\newcommand{\mrkd}{Mrk~1511}

\newcommand{\posteriorwidth}{6.7in}

\defcitealias{pancoast14b}{P14}
\defcitealias{Grier++17}{G17}

\newcommand{\iisarmax}{$36.9$}
\newcommand{\iisarmean}{$4.5_{-1.0}^{+1.4}$}
\newcommand{\iisarmedian}{$3.0_{-0.7}^{+1.1}$}
\newcommand{\iisarmin}{$0.96_{-0.40}^{+0.54}$}
\newcommand{\iisasigmar}{$4.4_{-1.1}^{+1.5}$}
\newcommand{\iisataumean}{$4.6_{-1.0}^{+1.4}$}
\newcommand{\iisataumedian}{$2.77_{-0.67}^{+0.92}$}
\newcommand{\iisabeta}{$1.23_{-0.21}^{+0.19}$}
\newcommand{\iisathetao}{$58_{-16}^{+25}$}
\newcommand{\iisathetai}{$44_{-13}^{+35}$}
\newcommand{\iisakappa}{$-0.17_{-0.17}^{+0.10}$}
\newcommand{\iisagamma}{$3.0_{-1.3}^{+1.3}$}
\newcommand{\iisaxi}{$0.70_{-0.33}^{+0.23}$}
\newcommand{\iisalogmbh}{$6.48_{-0.18}^{+0.21}$}
\newcommand{\iisafellip}{$<0.28$}
\newcommand{\iisafflow}{$0.75_{-0.18}^{+0.16}$}
\newcommand{\iisathetae}{$26_{-18}^{+26}$}
\newcommand{\iisasigmaturb}{$0.021_{-0.018}^{+0.049}$}
\newcommand{\iisainflowoutflow}{$0.68_{-0.32}^{+0.19}$}

\newcommand{\mrkarmax}{$40.4$}
\newcommand{\mrkarmean}{$8.23_{-0.53}^{+0.54}$}
\newcommand{\mrkarmedian}{$6.53_{-0.55}^{+0.57}$}
\newcommand{\mrkarmin}{$0.90_{-0.62}^{+0.81}$}
\newcommand{\mrkasigmar}{$6.41_{-0.57}^{+0.69}$}
\newcommand{\mrkataumean}{$7.43_{-0.41}^{+0.45}$}
\newcommand{\mrkataumedian}{$5.56_{-0.44}^{+0.43}$}
\newcommand{\mrkabeta}{$0.87_{-0.10}^{+0.13}$}
\newcommand{\mrkathetao}{$14.1_{-3.7}^{+4.8}$}
\newcommand{\mrkathetai}{$19.8_{-5.4}^{+6.0}$}
\newcommand{\mrkakappa}{$-0.03_{-0.10}^{+0.13}$}
\newcommand{\mrkagamma}{$3.9_{-1.3}^{+0.8}$}
\newcommand{\mrkaxi}{$0.15_{-0.10}^{+0.16}$}
\newcommand{\mrkalogmbh}{$7.50_{-0.18}^{+0.25}$}
\newcommand{\mrkafellip}{$0.51_{-0.15}^{+0.10}$}
\newcommand{\mrkafflow}{$0.75_{-0.17}^{+0.17}$}
\newcommand{\mrkathetae}{$17_{-12}^{+15}$}
\newcommand{\mrkasigmaturb}{$0.009_{-0.007}^{+0.022}$}
\newcommand{\mrkainflowoutflow}{$0.45_{-0.09}^{+0.13}$}

\newcommand{\mrkbrmax}{$36.3$}
\newcommand{\mrkbrmean}{$8.1_{-1.7}^{+1.8}$}
\newcommand{\mrkbrmedian}{$6.3_{-1.4}^{+1.5}$}
\newcommand{\mrkbrmin}{$2.08_{-0.85}^{+0.89}$}
\newcommand{\mrkbsigmar}{$6.1_{-1.6}^{+2.4}$}
\newcommand{\mrkbtaumean}{$7.5_{-1.6}^{+1.7}$}
\newcommand{\mrkbtaumedian}{$5.6_{-1.2}^{+1.2}$}
\newcommand{\mrkbbeta}{$1.02_{-0.17}^{+0.16}$}
\newcommand{\mrkbthetao}{$15.3_{-2.5}^{+3.9}$}
\newcommand{\mrkbthetai}{$26.0_{-4.3}^{+6.0}$}
\newcommand{\mrkbkappa}{$-0.224_{-0.078}^{+0.059}$}
\newcommand{\mrkbgamma}{$3.8_{-1.3}^{+0.8}$}
\newcommand{\mrkbxi}{$<0.071$}
\newcommand{\mrkblogmbh}{$7.46_{-0.21}^{+0.15}$}
\newcommand{\mrkbfellip}{$0.104_{-0.068}^{+0.082}$}
\newcommand{\mrkbfflow}{$0.75_{-0.18}^{+0.18}$}
\newcommand{\mrkbthetae}{$14_{-10}^{+16}$}
\newcommand{\mrkbsigmaturb}{$0.005_{-0.004}^{+0.011}$}
\newcommand{\mrkbinflowoutflow}{$0.838_{-0.092}^{+0.081}$}

\newcommand{\mrkcrmax}{$41.2$}
\newcommand{\mrkcrmean}{$13.3_{-1.3}^{+1.4}$}
\newcommand{\mrkcrmedian}{$12.2_{-1.5}^{+1.4}$}
\newcommand{\mrkcrmin}{$9.2_{-2.9}^{+2.1}$}
\newcommand{\mrkcsigmar}{$3.7_{-1.4}^{+3.2}$}
\newcommand{\mrkctaumean}{$11.8_{-1.2}^{+1.3}$}
\newcommand{\mrkctaumedian}{$11.2_{-1.3}^{+1.2}$}
\newcommand{\mrkcbeta}{$1.04_{-0.64}^{+0.71}$}
\newcommand{\mrkcthetao}{$41.0_{-4.1}^{+4.3}$}
\newcommand{\mrkcthetai}{$29.1_{-3.4}^{+3.4}$}
\newcommand{\mrkckappa}{$<-0.46$}
\newcommand{\mrkcgamma}{$3.2_{-1.2}^{+1.1}$}
\newcommand{\mrkcxi}{$<0.063$}
\newcommand{\mrkclogmbh}{$7.58_{-0.08}^{+0.08}$}
\newcommand{\mrkcfellip}{$<0.081$}
\newcommand{\mrkcfflow}{$0.76_{-0.17}^{+0.16}$}
\newcommand{\mrkcthetae}{$21.7_{-6.0}^{+7.8}$}
\newcommand{\mrkcsigmaturb}{$0.0037_{-0.0023}^{+0.0065}$}
\newcommand{\mrkcinflowoutflow}{$0.868_{-0.091}^{+0.053}$}

\newcommand{\mrkdrmax}{$40.4$}
\newcommand{\mrkdrmean}{$5.52_{-0.50}^{+0.55}$}
\newcommand{\mrkdrmedian}{$4.95_{-0.52}^{+0.56}$}
\newcommand{\mrkdrmin}{$0.72_{-0.53}^{+0.96}$}
\newcommand{\mrkdsigmar}{$2.85_{-0.40}^{+0.45}$}
\newcommand{\mrkdtaumean}{$5.94_{-0.46}^{+0.45}$}
\newcommand{\mrkdtaumedian}{$5.07_{-0.52}^{+0.50}$}
\newcommand{\mrkdbeta}{$0.61_{-0.10}^{+0.14}$}
\newcommand{\mrkdthetao}{$36_{-10}^{+9}$}
\newcommand{\mrkdthetai}{$19.3_{-4.7}^{+5.7}$}
\newcommand{\mrkdkappa}{$<-0.39$}
\newcommand{\mrkdgamma}{$<1.9$}
\newcommand{\mrkdxi}{$0.85_{-0.19}^{+0.09}$}
\newcommand{\mrkdlogmbh}{$7.11_{-0.17}^{+0.20}$}
\newcommand{\mrkdfellip}{$0.62_{-0.14}^{+0.16}$}
\newcommand{\mrkdfflow}{$0.27_{-0.18}^{+0.19}$}
\newcommand{\mrkdthetae}{$9_{-6}^{+14}$}
\newcommand{\mrkdsigmaturb}{$>0.029$}
\newcommand{\mrkdinflowoutflow}{$-0.35_{-0.14}^{+0.16}$}

\newcommand{\ngcarmax}{$20.3$}
\newcommand{\ngcarmean}{$3.41_{-0.55}^{+0.51}$}
\newcommand{\ngcarmedian}{$2.59_{-0.45}^{+0.52}$}
\newcommand{\ngcarmin}{$1.00_{-0.65}^{+0.80}$}
\newcommand{\ngcasigmar}{$2.41_{-0.49}^{+0.72}$}
\newcommand{\ngcataumean}{$3.29_{-0.40}^{+0.48}$}
\newcommand{\ngcataumedian}{$2.43_{-0.37}^{+0.42}$}
\newcommand{\ngcabeta}{$1.01_{-0.25}^{+0.61}$}
\newcommand{\ngcathetao}{$43_{-19}^{+22}$}
\newcommand{\ngcathetai}{$32_{-10}^{+19}$}
\newcommand{\ngcakappa}{$-0.25_{-0.22}^{+0.28}$}
\newcommand{\ngcagamma}{$<3.0$}
\newcommand{\ngcaxi}{$0.41_{-0.30}^{+0.27}$}
\newcommand{\ngcalogmbh}{$6.65_{-0.15}^{+0.27}$}
\newcommand{\ngcafellip}{$0.65_{-0.33}^{+0.15}$}
\newcommand{\ngcafflow}{$0.55_{-0.39}^{+0.31}$}
\newcommand{\ngcathetae}{$14_{-10}^{+24}$}
\newcommand{\ngcasigmaturb}{$>0.014$}
\newcommand{\ngcainflowoutflow}{$0.09_{-0.71}^{+0.22}$}

\newcommand{\zwiarmax}{$57.2$}
\newcommand{\zwiarmean}{$6.94_{-0.97}^{+0.99}$}
\newcommand{\zwiarmedian}{$4.59_{-0.78}^{+0.73}$}
\newcommand{\zwiarmin}{$2.19_{-0.61}^{+0.72}$}
\newcommand{\zwiasigmar}{$6.3_{-1.5}^{+1.9}$}
\newcommand{\zwiataumean}{$6.47_{-0.87}^{+0.90}$}
\newcommand{\zwiataumedian}{$4.12_{-0.65}^{+0.69}$}
\newcommand{\zwiabeta}{$1.36_{-0.30}^{+0.31}$}
\newcommand{\zwiathetao}{$33.5_{-6.2}^{+6.4}$}
\newcommand{\zwiathetai}{$32.9_{-5.2}^{+6.1}$}
\newcommand{\zwiakappa}{$-0.417_{-0.054}^{+0.065}$}
\newcommand{\zwiagamma}{$3.3_{-1.3}^{+1.2}$}
\newcommand{\zwiaxi}{$<0.080$}
\newcommand{\zwialogmbh}{$6.94_{-0.14}^{+0.14}$}
\newcommand{\zwiafellip}{$<0.15$}
\newcommand{\zwiafflow}{$0.74_{-0.18}^{+0.17}$}
\newcommand{\zwiathetae}{$10.9_{-7.2}^{+9.0}$}
\newcommand{\zwiasigmaturb}{$0.024_{-0.021}^{+0.048}$}
\newcommand{\zwiainflowoutflow}{$0.88_{-0.12}^{+0.07}$}

\newcommand{\combinedmeanfrmssigma}{$0.57 \pm 0.07$}
\newcommand{\combinedsigmafrmssigma}{$0.14 \pm 0.10$}
\newcommand{\combinedpredfrmssigma}{$0.57 \pm 0.19$}
\newcommand{\combinedmeanfmeansigma}{$0.43 \pm 0.09$}
\newcommand{\combinedsigmafmeansigma}{$0.22 \pm 0.11$}
\newcommand{\combinedpredfmeansigma}{$0.43 \pm 0.26$}
\newcommand{\combinedmeanfmeanfwhm}{$0.00 \pm 0.14$}
\newcommand{\combinedsigmafmeanfwhm}{$0.46 \pm 0.12$}
\newcommand{\combinedpredfmeanfwhm}{$0.00 \pm 0.50$}
\newcommand{\lampmeanfrmssigma}{$0.60 \pm 0.16$}
\newcommand{\lampsigmafrmssigma}{$0.22 \pm 0.19$}
\newcommand{\lamppredfrmssigma}{$0.60 \pm 0.32$}
\newcommand{\lampmeanfmeansigma}{$0.47 \pm 0.15$}
\newcommand{\lampsigmafmeansigma}{$0.25 \pm 0.17$}
\newcommand{\lamppredfmeansigma}{$0.47 \pm 0.34$}
\newcommand{\lampmeanfmeanfwhm}{$-0.21 \pm 0.14$}
\newcommand{\lampsigmafmeanfwhm}{$0.23 \pm 0.16$}
\newcommand{\lamppredfmeanfwhm}{$-0.21 \pm 0.31$}

\shorttitle{Modeling the BLR in LAMP 2011 AGN} 
\shortauthors{Williams et al.}
 
\begin{document}

\title{The Lick AGN Monitoring Project 2011: Dynamical Modeling of the Broad-Line Region}

\author[0000-0002-4645-6578]{Peter R. Williams}
\affiliation{Department of Physics and Astronomy, University of California, Los Angeles, CA 90095-1547, USA}

\author[0000-0003-1065-5046]{Anna Pancoast}
\altaffiliation{Einstein Fellow}
\affiliation{Harvard-Smithsonian Center for Astrophysics, 60 Garden Street, Cambridge, MA 02138, USA}

\author[0000-0002-8460-0390]{Tommaso Treu}
\affiliation{Department of Physics and Astronomy, University of California, Los Angeles, CA 90095-1547, USA}

\author{Brendon J. Brewer}
\affiliation{Department of Statistics, The University of Auckland, Private Bag 92019, Auckland 1142, New Zealand}

\author[0000-0002-3026-0562]{Aaron J. Barth}
\affiliation{Department of Physics and Astronomy, 4129 Frederick Reines Hall, University of California, Irvine, CA, 92697-4575, USA}

\author[0000-0003-2064-0518]{Vardha N. Bennert}
\affiliation{Physics Department, California Polytechnic State University San Luis Obispo, CA 93407, USA}

\author{Tabitha Buehler}
\affiliation{Department of Physics and Astronomy, N283 ESC, Brigham Young University, Provo, UT 84602-4360, USA}
\affiliation{Department of Physics and Astronomy, University of Utah, Salt Lake City, UT 84112, USA}

\author{Gabriela Canalizo}
\affiliation{Department of Physics and Astronomy, University of California, Riverside, CA 92521, USA}

\author[0000-0003-1673-970X]{S. Bradley Cenko}
\affiliation{Astrophysics Science Division, NASA Goddard Space Flight Center, Mail Code 661, Greenbelt, MD 20771, USA}
\affiliation{Joint Space-Science Institute, University of Maryland, College Park, MD 20742, USA}

\author{Kelsey I. Clubb}
\affiliation{Department of Astronomy, University of California, Berkeley, CA 94720-3411, USA}

\author[0000-0003-1371-6019]{Michael C. Cooper}
\affiliation{Department of Physics and Astronomy, 4129 Frederick Reines Hall, University of California, Irvine, CA, 92697-4575, USA}

\author{Alexei V. Filippenko}
\affiliation{Department of Astronomy, University of California, Berkeley, CA 94720-3411, USA}
\affiliation{Miller Senior Fellow, Miller Institute for Basic Research in Science, University of
California, Berkeley, CA  94720}

\author[0000-0002-3739-0423]{Elinor Gates}
\affiliation{Lick Observatory, P.O. Box 85, Mount Hamilton, CA 95140, USA}

\author[0000-0002-6353-1111]{Sebastian F. Hoenig}
\affiliation{Department of Physics \& Astronomy, University of Southampton, Southampton, SO17 1BJ, United Kingdom}

\author[0000-0003-0634-8449]{Michael D. Joner}
\affiliation{Department of Physics and Astronomy, N283 ESC, Brigham Young University, Provo, UT 84602-4360, USA}

\author{Michael T. Kandrashoff}
\affiliation{Department of Astronomy, University of California, Berkeley, CA 94720-3411, USA}

\author{Clifton David Laney}
\affiliation{Department of Physics and Astronomy, Western Kentucky University, 1906 College Heights Boulevard, Bowling Green, KY 42101, USA}

\author{Mariana S. Lazarova}
\affiliation{University of Northern Colorado, Greeley, CO 80639, USA}

\author{Weidong Li}
\altaffiliation{Deceased 2011 December 12.}
\affiliation{Department of Astronomy, University of California, Berkeley, CA 94720-3411, USA}

\author[0000-0001-6919-1237]{Matthew A. Malkan}
\affiliation{Department of Physics and Astronomy, University of California, Los Angeles, CA 90095-1547, USA}

\author{Jacob Rex}
\affiliation{Department of Astronomy, University of California, Berkeley, CA 94720-3411, USA}

\author{Jeffrey M. Silverman}
\affiliation{Department of Astronomy, University of California, Berkeley, CA 94720-3411, USA}
\affiliation{Samba TV, San Francisco, CA 94107, USA}

\author[0000-0002-9599-310X]{Erik Tollerud}
\affiliation{Space Telescope Science Institute}

\author[0000-0002-1881-5908]{Jonelle L. Walsh}
\affiliation{George P. and Cynthia Woods Mitchell Institute for Fundamental Physics and Astronomy, Department of Physics and Astronomy, Texas A\&M University, College Station, TX 77843-4242, USA}

\author{Jong-Hak Woo}
\affiliation{Physics and Astronomy Department, Seoul National University, Seoul Korea, 08826}

\correspondingauthor{Peter R. Williams}
\email{pwilliams@astro.ucla.edu}

\begin{abstract}
We present models of the \Hb-emitting broad-line region (BLR) in seven Seyfert 1 galaxies from the Lick AGN (Active Galactic Nucleus) Monitoring Project 2011 sample, drawing inferences on the BLR structure and dynamics as well as the mass of the central supermassive black hole.
We find that the BLR is generally a thick disk, viewed close to face-on, with preferential emission back toward the ionizing source.
The dynamics in our sample range from near-circular elliptical orbits to inflowing or outflowing trajectories.
We measure black hole masses of $\log_{10}(M_{\rm BH}/M_\odot) = $ \iisalogmbh\ for \iisa, \mrkalogmbh\ for \mrka, \mrkblogmbh\ for \mrkb, \mrkclogmbh\ for \mrkc, \mrkdlogmbh\ for \mrkd, \ngcalogmbh\ for \ngca, and \zwialogmbh\ for \zwia.
We use these black hole mass measurements along with cross-correlation time lags and line widths to recover the scale factor $f$ used in traditional reverberation mapping measurements.
Combining our results with other studies that use this modeling technique, bringing our sample size to 16, we calculate a scale factor that can be used for measuring black hole masses in other reverberation mapping campaigns.
When using the root-mean-square (rms) spectrum and using the line dispersion to measure the line width, we find \logfrmssigma$_{\rm pred} = $ \combinedpredfrmssigma.
Finally, we search for correlations between $f$ and other AGN and BLR parameters and find marginal evidence that $f$ is correlated with $M_{\rm BH}$ and the BLR inclination angle, but no significant evidence of a correlation with the AGN luminosity or Eddington ratio.
\end{abstract}



\section{Introduction}
\label{sect:intro}

Supermassive black holes are thought to play an important role in galaxy formation and evolution.
Tight correlations in the local universe between black hole masses and host-galaxy properties \citep[e.g.,][]{Magorrian++98,ferrarese00,gebhardt00} suggest a fundamental link between the growth of black holes and their hosts.
Depending on the relative timing of black hole and host-galaxy growth, one might expect an evolution of these scaling relations, so accurate measurements of black hole masses across cosmic time are essential for testing the predictions of different evolutionary scenarios.

\begin{deluxetable*}{llcccccccc}
\tablecaption{AGN and data properties}
\tablewidth{0pt}
\tablehead{ 
\colhead{Galaxy} & 
\colhead{Alt. Name} &
\colhead{$z$} &
\colhead{$N_{\rm spec}$}  &
\colhead{$N_{\rm phot}$}  &
\colhead{$(\Delta t)_{\rm spec}$}  &
\colhead{$(\Delta t)_{\rm phot}$}  &
\colhead{$(S/N)/{\rm pix}$} &
\colhead{$(S/N)/{\rm pix}$} &
\colhead{$(S/N)/{\rm pix}$}\\
\colhead{} & 
\colhead{} & 
\colhead{} & 
\colhead{}  &
\colhead{} & 
\colhead{(days)}  &
\colhead{(days)} &
\colhead{\kov} &
\colhead{\bg} &
\colhead{\veron}
}
\startdata
\mrka	&	 	&	0.0234	&	55	&	170	&	2.6	&	0.9	&	19.1	&	19.2	&	19.8	\\
\mrkb	&	 	&	0.0417	&	36	&	93	&	1.7	&	1.1	&	14.5	&	14.1	&	14.6	\\
\mrkc	&	 PG 1351$+$695 	&	0.0305	&	34	&	64	&	2.3	&	1.7	&	17.9	&	18.3	&	18.2	\\
\mrkd	&	 NGC 5940 	&	0.0339	&	40	&	71	&	1.9	&	1.5	&	22.6	&	21.7	&	22.4	\\
\ngca	&	 Mrk 1330 	&	0.0090	&	43	&	75	&	1.8	&	1.1	&	27.9	&	28.0	&	29.2	\\
\zwia	&	 	&	0.0279	&	29	&	69	&	3.2	&	2.4	&	5.7	&	5.7	&	5.3	\\
\iisa	&	II SZ 10	&	0.0343	&	35	&	63	&	2.2	&	1.6	&	21.4	&	21.5	&	20.3
\enddata
\tablecomments{Properties and observing information for the seven AGN modeled in this work.
The redshifts $z$ are from \citet{barth15}.
$N_{\rm spec}$ and $N_{\rm phot}$ are the number of spectroscopic and photometric observations, respectively.
The columns $(\Delta t)_{\rm spec}$ and $(\Delta t)_{\rm phot}$ give the average spacing between subsequent spectroscopic and photometric observations, respectively.
The median spacing between subsequent observations for all AGNs was one day.
$(S/N)/{\rm pix}$ is the median signal to noise per pixel in the \Hb\ spectrum from spectral decomposition using the three \feii\ templates discussed in Section \ref{sect:decomp}.
\label{table_agnproperties}}
\end{deluxetable*} 

In nearby galaxies, black hole masses can be measured through stellar or gas kinematics within the black hole sphere of influence \citep[e.g.,][]{kormendy95,ferrarese05}, but this is not possible at distances greater than $\sim$100 Mpc where even the largest black holes' spheres of influence cannot be resolved.
The technique of reverberation mapping \citep{blandford82,peterson93} substitutes time resolution for spatial resolution by measuring the response of broad emission lines to active galactic nucleus (AGN) continuum variations, enabling measurements out to cosmological distances.
The time lag $\tau$ between continuum and emission-line variations can be combined with the speed of light to obtain a characteristic radius of the BLR, while the line width measures the velocity $v$ of the emitting gas.
By assuming that the motion of the gas in the broad-line region (BLR) is dominated by the black hole's gravity, one can make a virial estimate of the black hole's mass,
\begin{align}
    M_{\rm BH} = f \frac{c\tau v^2}{G},
\end{align}
\label{eq:mbheq}
\noindent
where $f$ is a scale factor of order unity that accounts for the detailed structure, orientation, and dynamics of the BLR.
Typically, an average value of $f$ is used, found by aligning AGNs with the $M_{\rm BH} - \sigma_*$ relation for quiescent galaxies \citep[e.g.,][]{onken04,collin06,woo10,woo13,graham11,grier13b,Batiste++17}.
The scatter in the $M_{\rm BH} - \sigma_*$ relation introduces an uncertainty of $\sim$0.4 dex for individual $M_{\rm BH}$ measurements \citep{park12b}, making it the largest source of uncertainty in reverberation mapping $M_{\rm BH}$ measurements.
It is therefore very important to understand reverberation mapping results and the $f$ factor; they calibrate all the single-epoch black hole masses throughout the Universe \citep[e.g.,][]{Shen+10}.

Since there are multiple ways to measure the line width, more than one version of $f$ exists.
Typically, either the line dispersion ($\sigma_{\rm line}$, second central moment of the emission line profile) is measured using the root-mean-square (rms) spectrum, or the full width at half-maximum intensity (FWHM) is measured in the time-averaged spectrum.
In cases where the rms spectrum is unavailable, such as in single-epoch measurements, the line dispersion measured in the time-averaged spectrum is often used.
For clarity, we will specify which $f$ we are discussing by using the notation  $f_{s,v}$, where $s$ is the spectrum used (mean or rms) and $v$ is the type of line width (FWHM or $\sigma_{\rm line}$).

The cross-correlation and single-epoch techniques yield a BLR size, but they do not provide information about the gas structure or dynamics needed to determine $f$ for an individual AGN.
Recently, high-quality reverberation mapping datasets have enabled velocity-resolved analyses that look individually at how different parts of the broad emission line change, allowing inferences to be drawn about the structure and dynamics of the BLR \citep{bentz09,denney09,denney10,barth11,barth11b,grier13a,Du++16,Pei++17}.
These results are generally consistent with inflowing gas or elliptical orbits, but some have shown signs of gas outflow \citep{Denney++09,Du++16}.
Other studies have used the code MEMECHO \citep{horne91,horne94} to recover the two-dimensional (2D) transfer function, which defines how continuum changes map to broad-line flux variations as a function of line-of-sight velocity and time delay \citep{bentz10b,grier13a,skielboe15}.
On their own, the resulting maps do not provide details of the BLR structure and kinematics, but can be compared to the transfer functions that result from specific BLR models.

Recent efforts have aimed to measure $M_{\rm BH}$ independent of $f$ by modeling the structure and dynamics of the BLR directly \citep{brewer11b,pancoast11,pancoast12,pancoast14a,li13}.
\citet[][hereafter \citetalias{pancoast14b}]{pancoast14b} used the model of \citet{pancoast14a} to model the BLR of five AGNs in the Lick AGN Monitoring Project 2008 sample \citep[LAMP 2008,][]{walsh09,bentz09}, and \citet[][hereafter \citetalias{Grier++17}]{Grier++17} expanded the sample by modeling four AGNs from a 2010 campaign carried out at MDM Observatory.
The results from these analyses find an \Hb-emitting BLR that is a thick disk with kinematics that are best described by a combination of elliptical orbits and inflowing gas, consistent with the velocity-resolved reverberation mapping methods.
They also measure scale factors for all the AGNs in their samples and find a mean scale factor of \meanlogfrmssigma$ = 0.54 \pm 0.17$ \citepalias{Grier++17}, which is consistent with previous measurements made using the $M_{\rm BH} - \sigma_*$ relation.

In this paper, we expand the sample of AGN modeled using the techniques of \citet{pancoast14a} from 9 to 16 by analyzing the data for seven of the AGNs from the Lick AGN Monitoring Project 2011 campaign \citep[LAMP 2011,][]{barth15}.
This nearly doubles the sample of AGNs analyzed using this method and will help uncover general trends in BLR properties.
We also aim to gain a better understanding of how $f$ is related to other AGN and BLR properties.
Since $f$ measurements so far are mostly based on local low-luminosity Seyfert galaxies, understanding how $f$ depends on (for example) continuum luminosity and Eddington ratio will help reduce uncertainties when reverberation mapping techniques are extrapolated and applied to AGNs across the entire Universe.
In Section \ref{sect:data}, we describe the spectroscopic and photometric monitoring data and discuss the spectral decomposition method of extracting \Hb\ from the rest of the AGN spectrum.
Section \ref{sect:model} summarizes the geometrical and dynamical model from \citet{pancoast14a} that we used to model the BLR.
We discuss in Section \ref{sect:results} the modeling results for each individual AGN in our sample.
In Section \ref{sect:discussion}, we combine our results with those of \citetalias{pancoast14b} and \citetalias{Grier++17} to calculate a mean scale factor $\bar{f}$ and look for useful correlations between $f$ and other parameters.
Finally, we conclude in Section \ref{sect:summary}.


\section{Data}
\label{sect:data}

The data used in this paper were taken as part of the Lick AGN Monitoring Project 2011 campaign \citep[LAMP 2011;][]{barth15}.
Photometric monitoring of the AGNs was carried out in the Johnson $V$ band using several telescopes:
the 0.76 m Katzman Automatic Imaging Telescope (KAIT) at Lick Observatory \citep{filippenko01}; the 0.91 m telescope at West Mountain Observatory (WMO); the 2 m Faulkes Telescope North at Mt. Haleakala Hawaii and the Faulkes Telescope South at Siding Spring Australia, both part of the Las Cumbres Observatory network \citep[LCO,][]{Brown++13}; the 0.60 m Super-LOTIS telescope at the Steward Observatory, Kitt Peak; and the Palomar 1.5 m telescope at Palomar Observatory \citep{cenko06}.

Spectra were obtained over the course of 69 nights from 2011 March 27 to June 13 with the Kast double spectrograph mounted on the Shane 3 m telescope at Lick Observatory.
Owing to poor weather, a substantial fraction of the nights were lost.
This analysis only uses the spectra from the blue side of the Kast spectrograph, which covered 3440--5515 \AA\ at 1.02 \AA\ per pixel.
The spectra were calibrated between nights using the procedure of \citet{vanGron+92}, assuming the flux of the \oiii\ doublet remained constant throughout the campaign.
In addition to the LAMP 2011 observations, \mrka\ received twelve additional observations from January through March 2011 and \zwia\ received three additional observations after the campaign in order to extend the light curve.
All additional observations were also taken with the Kast double spectrograph.

\begin{figure}[h!]
\begin{center}
\includegraphics[width=3in]{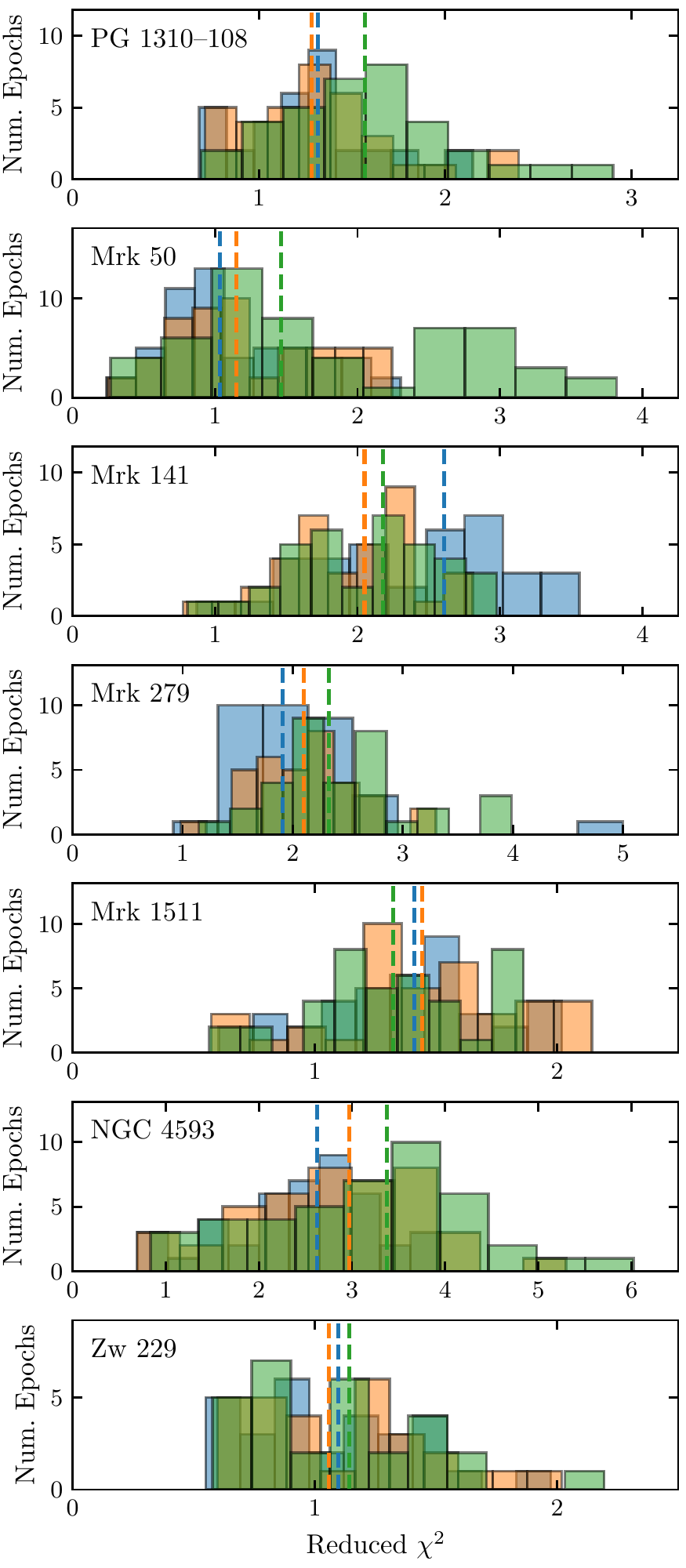}  
\caption{Distribution of the reduced $\chi^2$ values for fits to the spectra using the \kov\ (blue), \bg\ (orange), and \veron\ (green) \feii\ templates.
The vertical dashed lines indicate the median reduced $\chi^2$ value.
\label{fig:chisq}}
\end{center}
\end{figure}

In total, 15 AGNs were observed during the campaign, but only 7 had sufficient data quality and continuum and \Hb\ variations for the analysis in this paper.
General properties of the targets analyzed in this paper along with information on their observations are given in Table \ref{table_agnproperties}.

\begin{figure*}[h!]
\begin{center}
\includegraphics[width=6.5in]{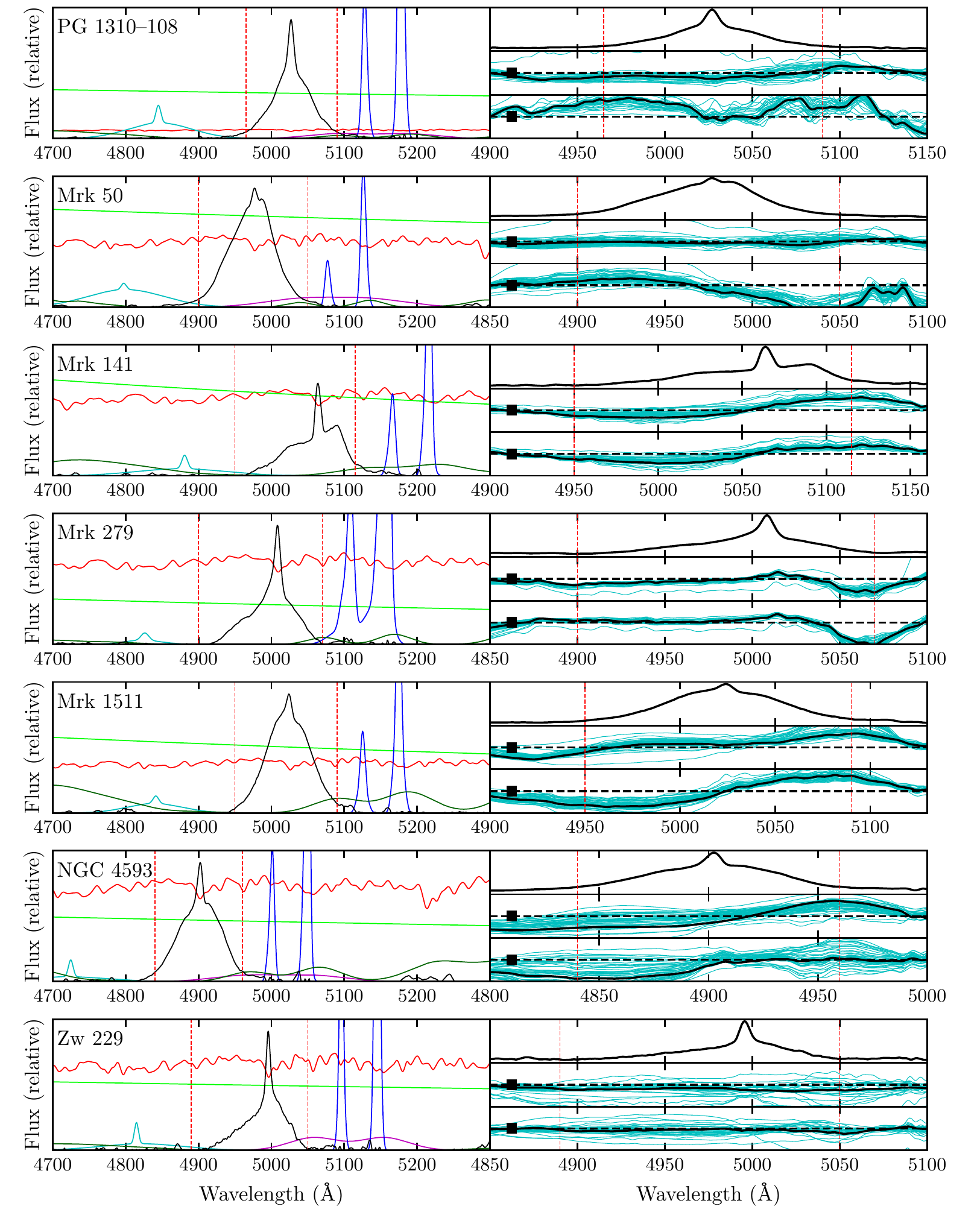}  
\caption{\textit{Left}: Spectral decomposition using the \kov\ \feii\ template.
The components shown are starlight in red, the AGN featureless continuum in lime green, \hei\ in magenta, \heii\ in cyan, \oiii\ in blue, \feii\ in dark green, and the residual \Hb\ in black.
The vertical dashed red lines indicate the wavelength range that was used when fitting the BLR model.
\textit{Right-top}: The mean \Hb\ profile shape, for reference in the right-middle and right-bottom panels.
\textit{Right-middle}: In black is the mean \Hb\ spectrum found using the \kov\ \feii\ template minus the mean \Hb\ spectrum found using the \bg\ template.
The cyan lines show the same thing, but for every observational epoch.
The black bar on the left shows the mean uncertainty in the spectra over the modeled wavelength range.
\textit{Right-bottom}: Same as right-middle, but for the \kov\ and \veron\ templates.
For the full spectral decompositions, including the full model fits and residual spectra, see \citet{barth15}.
\label{fig:decomps}}
\end{center}
\end{figure*}

\subsection{Spectral Decomposition}
\label{sect:decomp}
When we fit BLR models to the data, we not only allow for variations in the total \Hb\ flux, but also variations in the detailed shape of the \Hb\ broad emission-line spectrum.
Because of this, it is critical to disentangle the \Hb\ emission from other features contributing to the AGN spectrum.
In particular, features such as \hei\ or \feii\ emission that preferentially affect the red wing of the \Hb\ profile must be properly subtracted, otherwise the models will attempt to fit an asymmetry that is not intrinsic to the broad \Hb\ line.
In order to accurately isolate the \Hb\ profile, we fit for contributions in the vicinity of \Hb\ that may strongly affect its shape --- AGN continuum; host-galaxy starlight; \oiii~$\lambda 4959$ and $\lambda 5007$; \heii~$\lambda 4686$; \hei~$\lambda 4471$, $\lambda 4922$, and $\lambda 5016$; and \feii~emission blends.
These components were fit to the spectra by minimizing $\chi^2$ with the Levenberg-Marquardt routines in the IDL package {\sc mpfit} \citep{Markwardt09}.
After subtracting these features from the data, we are left with the residual \Hb\ spectrum.
The full details of how the spectra were decomposed into their individual components are given by \citet{barth15}.

The full fitting procedure was carried out three times for each spectrum using three different \feii\ templates from \citet{boroson92}, \citet{kovacevic10}, and \citet{veron-cetty04}, hereafter \bg, \kov, \veron\ (respectively).
In addition to free parameters for the velocity shift and broadening kernel, the \bg\ and \veron\ templates each have one free parameter describing the flux normalization.
The \kov\ template has five components and so has five normalization parameters, making it more flexible, in general.
For each AGN, we show distributions of the reduced $\chi^2$ values from fitting each epoch in Figure \ref{fig:chisq}.
The reduced $\chi^2$ values do not include any systematic uncertainties from (for example) flux calibration, but they allow for a relative comparison between the templates in order to determine which provide the best fit to the data.
Generally, the distributions of reduced $\chi^2$ are similar for each template, which we take to mean that each decomposition is equally valid.
We choose to run the dynamical modeling procedure using the spectra from all three decompositions and combine the resulting model parameter posterior samples, weighting each run equally.
An exception to this is \mrka, for which the \veron\ template produced poor fits for some epochs.
For this object, we adopt only the decompositions that use the \kov\ and \bg\ templates.

In Figure \ref{fig:decomps}, we show the results of the spectral decomposition for each AGN.
The left panels show the individual components for the mean spectrum, found using the \kov\ \feii\ template.
We note that there is a degeneracy between the \hei\ $\lambda 4922$ and $\lambda 5016$ \AA\ lines and two features in the \feii\ templates, which appears in the fit for \zwia.
However, as discussed by \citet{barth15}, the sum of these two features is well determined, so the residual \Hb\ profile is robust.
On the right, we show the mean \Hb\ profile derived with the \kov\ \feii\ template as well as the difference in \Hb\ profiles from using the other \feii\ templates.
The thick black bar shows the mean uncertainty in the spectra across the modeled wavelength range.
The difference between templates exceeds the flux uncertainty for some AGNs, indicating that template choice is an important factor that may influence the modeling results.
In particular, \mrkd\ and \ngca\ both have prominent \feii\ emission and therefore the \feii\ fits inherently have a strong effect on the resulting \Hb\ profile.
For objects such as Mrk 141, Mrk 1511, and NGC 4593, the \kov\ template fits give a mean \Hb\ profile with a stronger red wing and weaker blue wing than the \bg\ and \veron\ template fits.
Asymmetries in the line profile are caused by asymmetries in the BLR properties, so we might expect discrepancies in the inferred model asymmetry parameters for these objects.


\section{The Geometric and Dynamical Model of the BLR}
\label{sect:model}

We use a simply parameterized phenomenological model of the BLR, described by \citet{pancoast14a}, to model the AGN data sets.
In this framework, the \Hb-emitting BLR is modeled as a distribution of point particles around a central ionizing continuum source that we take to be point-like in nature and isotropically emitting.
Each particle receives the continuum emission after a time lag determined by its position, and then is assumed to instantaneously reprocess the light and re-emit it in the direction of the observer.
The wavelength of the re-emitted light is centered on \Hb, with a Doppler shift determined by the particle's velocity.
By feeding a continuum light curve through the model, we can produce a time-series of spectra that we can then directly compare to data.

\begin{deluxetable}{ll}
\tablecaption{Model parameters and priors}
\tablewidth{0pt}
\tablehead{ 
\colhead{Parameter} & 
\colhead{Prior}
}
\startdata
$\mu$                           & LogUniform($1.02\times 10^{-3}$ light days, $\Delta t_{\rm data}$) \\
$\beta$                         & Uniform(0,2) \\
$F$                             & Uniform(0,1) \\
$\theta_i$                      & Uniform($\cos\theta_i(0,\pi/2)$) \\
$\theta_o$                      & Uniform(0,$\pi/2$) \\
$\kappa$                        & Uniform(-0.5,0.5) \\
$\gamma$                        & Uniform(1,5) \\
$\xi$                           & Uniform(0,1) \\
$M_{\rm BH}$                    & LogUniform($2.78\times 10^4$,$1.67\times 10^9$ M$_\odot$) \\
$f_{\rm ellip}$                 & Uniform(0,1) \\
$f_{\rm flow}$                  & Uniform(0,1) \\
$\sigma_{\rho,{\rm circ}}$      & LogUniform(0.001,0.1) \\
$\sigma_{\Theta,{\rm circ}}$    & LogUniform(0.001,0.1) \\
$\sigma_{\rho,{\rm radial}}$    & LogUniform(0.001,0.1) \\
$\sigma_{\Theta,{\rm radial}}$  & LogUniform(0.001,0.1) \\
$\sigma_{\rm turb}$             & LogUniform(0.001,0.1) \\
$\theta_e$                      & Uniform(0,$\pi/2$)
\enddata
\tablecomments{List of BLR model parameters and their corresponding priors.
\label{table_priors}}
\end{deluxetable}

In order to calculate the line emission of a given BLR geometry at arbitrary times, we need to know the AGN continuum flux at arbitrary times before the time of emission.
To determine the continuum flux between data points, we model the AGN continuum using Gaussian processes, which has been shown to be a sufficiently good model for AGN light curves \citep{kelly09,kozlowski10,kozlowski16,macleod10,zu11,zu13}.
As discussed by \citet{skielboe15}, with the use of more general descriptions of the driving light curves, our inferences are robust with respect to the assumption of Gaussian processes, since Gaussian processes are effectively used as a flexible interpolator.
This has the advantage of allowing us to include the uncertainties of the continuum modeling in the uncertainties of our model parameters.
Additionally, we can extrapolate the continuum light curve to times before and after continuum monitoring in order to model the BLR response for the full extent of the spectroscopic monitoring campaign.

The full details of the BLR model and its limitations are discussed by \citet{pancoast14a}, but we summarize the main components below.

\subsection{Geometry}

We first assign radial positions to each particle drawn from a Gamma distribution which has a probability density function
\begin{eqnarray}
p(x|\alpha, \theta)  \propto x^{\alpha-1} \exp \left( - \frac{x}{\theta}   \right),
\end{eqnarray}
where $\alpha$ is the shape parameter and $\theta$ is the scale parameter.
We set a minimum radius of the BLR by shifting the distribution from the origin by the Schwarzschild radius $R_s = 2GM_{\rm BH}/c^2$ plus a free parameter $r_{\rm min}$.
We also set a maximum BLR radius $r_{\rm out} = c\Delta t_{\rm data}/2$, where $\Delta t_{\rm data}$ is the time between the first modeled point of the AGN continuum light curve and the first spectrum of the broad emission line.
This comes from the assumption that our observational campaign is sufficiently long to measure the response of the whole BLR.
We then perform a change of variables from $(\alpha, \theta, r_{\rm min})$ to
$(\mu, \beta, F)$:
\begin{eqnarray}
\mu &=& r_{\rm min} + \alpha \theta, \\
\beta &=& \frac{1}{\sqrt{\alpha}}, \\
F &=&  \frac{r_{\rm min}}{r_{\rm min} + \alpha \theta},
\end{eqnarray}
\noindent
where $\mu$ is the mean radius, $\beta$ determines the shape of the Gamma distribution, and $F$ is the minimum radius in units of $\mu$.
In this framework, the standard deviation of the radial distribution is given by $\sigma_r = (1-F)\mu \beta$.
The distribution of particles is then ``puffed up'' out of a plane by opening angle $\theta_o$ such that $\theta_o = 0^\circ$ corresponds to a flat disk and $\theta_o = 90^\circ$ corresponds to a sphere, and the plane of the distribution is inclined by an angle $\theta_i$ relative to the observer where $\theta_i = 0^\circ$ is face-on and $\theta_i = 90^\circ$ is edge-on.

The relative emission from each particle is weighted by a parameter 
\begin{eqnarray}
W(\phi) = \frac{1}{2} + \kappa \cos(\phi)
\end{eqnarray}
which allows for BLR asymmetry.
The angle $\phi$ is measured from the particle to the origin to the observer, and $\kappa$ is a free parameter between $-0.5$ and 0.5.
In this setup, $\kappa \rightarrow 0.5$ corresponds to emission from the near side of the BLR and $\kappa \rightarrow -0.5$ corresponds to emission from the far side.
These cases can physically be interpreted as gas that preferentially re-emits away from or back toward the ionizing source, respectively.
The broad line emission is allowed to preferentially come from the faces of the disk according to a parameter $\gamma$, which has a uniform prior between 1 and 5.
The angle between a point particle and the disk is
\begin{eqnarray}
\theta = {\rm acos} (\cos \theta_o + (1- \cos \theta_o) U^{\gamma} ),
\end{eqnarray}
where $U$ is drawn randomly from a uniform distribution between 0 and 1.
When $\gamma\rightarrow 1$, point particles are evenly distributed, and when $\gamma\rightarrow 5$, particles are clustered near the faces of the disk.
Finally, the accretion disk is allowed to be transparent to opaque according to the parameter $\xi$, ranging from 0 to 1.
When $\xi\rightarrow 0$, the midplane is opaque, and when $\xi\rightarrow 1$, the midplane is transparent.

\subsection{Dynamics}
The particle velocities are assigned based on the mass of the black hole, their radial position, and the parameters $f_{\rm ellip}$, $f_{\rm flow}$, $\theta_e$, and $\sigma_{\rm turb}$.
First, each particle is assigned to have its radial and tangential velocities drawn from a distribution centered either around the circular velocity or around the radial escape velocity.
The fraction of particles that are assigned near-circular orbits is given by $f_{\rm ellip}$, which has a uniform prior between 0 and 1.
The specific radial and tangential velocities of these particles are drawn from a Gaussian distribution centered on the circular velocity in the $v_r-v_\phi$ plane, with standard deviations $\sigma_{\rho,{\rm circ}}$ and $\sigma_{\Theta,{\rm circ}}$.

The remaining particles are then assigned to be either inflowing or outflowing according to a binary parameter $f_{\rm flow}$, where $0 < f_{\rm flow} < 0.5$ corresponds to inflow and $0.5 < f_{\rm flow} < 1$ corresponds to outflow.
The specific radial and tangential velocties for the inflowing and outflowing particles are drawn from Gaussian distributions centered on the inflowing and outflowing escape velocities (respectively) in the $v_r-v_\phi$ plane, with standard deviations $\sigma_{\rho,{\rm radial}}$ and $\sigma_{\Theta,{\rm radial}}$.
The angle $\theta_e$ then rotates the centers of these two distributions toward the circular orbit by an angle between $0^\circ$ and $90^\circ$ to allow for further flexibility.
This means that as $\theta_e\rightarrow 90^\circ$, all particles approach a distribution centered on the circular velocity, regardless of the value of $f_{\rm ellip}$.

Finally, the $\sigma_{\rm turb}$ parameter allows for random macroturbulent velocities according to
\begin{eqnarray}
v_{\rm turb} = \mathcal{N}(0, \sigma_{\rm turb})|v_{\rm circ}|,
\end{eqnarray}
where $\mathcal{N}(0, \sigma_{\rm turb})$ is a Gaussian distribution with mean 0 and standard deviation $\sigma_{\rm turb}$, and $v_{\rm turb}$ is added to the line-of-sight velocity.
The parameter $\sigma_{\rm turb}$ has a log-uniform prior between 0.001 and 0.1.

\subsection{Producing Emission-Line Spectra and Comparing to Data}
Given the continuum light-curve model and a model of the BLR, we can generate \Hb\ emission-line spectra at arbitrary times.
For each particle, we use the position and line-of-sight velocity to calculate the Doppler and gravitational redshifts and then use the strength of the continuum and the particle's emissivity properties to calculate the amount of flux contribution from that particle.
After combining the contributions from all particles, we blur the spectrum by the resolution of the instrument, $\Delta \lambda_{\rm dis}$, which is calibrated by comparing the width of the observed \oiii\,$\lambda 5007$ emission line from spectral decomposition, $\Delta \lambda_{\rm obs}$, to the intrinsic line width, $\Delta \lambda_{\rm true}$, taken from \citet{whittle92}:
\begin{equation}
    \Delta \lambda_{\rm dis}^2 \approx \Delta \lambda_{\rm obs}^2 - \Delta \lambda_{\rm true}^2.
\end{equation}
Finally, the AGN redshift is left as a free parameter with a Gaussian prior having standard deviation 0.25--0.5~\AA, depending on the AGN.

We use a Gaussian likelihood function to compare the observed time-series of \Hb\ spectra from the spectral decomposition to the spectra produced by the model.
To explore the parameter space of the continuum light-curve model and BLR model, we use the diffusive nested sampling code {\sc DNest3} \citep{brewer11}.
In addition to producing posterior probability density functions (PDFs), diffusive nested sampling also calculates the normalization term, the ``evidence,'' which allows for model comparison.

In practice, our simplified BLR model is unable to reproduce all of the details of the emission line and its fluctuations to within the small spectral uncertainties.
We account for this systematic uncertainty by softening the likelihood function with a ``temperature'' $T$, where $T \ge 1$.
We divide the log of the likelihood by $T$, which is equivalent to multiplying the spectral uncertainties by $\sqrt{T}$ in the case of a Gaussian likelihood function.
When the temperature is too small, the continuum model hyperparameters are overfitted or the model is unable to efficiently explore parameter space.
We choose the lowest temperature for which this is not the case.
In our sample, we use $T = 150$--180 for \mrka, $T = 30$--40 for \mrkb, $T = 60$--140 for \mrkc, $T = 20$--25 for \mrkd, $T = 35$ for \iisa, $T = 150$--200 for \ngca, and $T = 70$ for \zwia.
We test the convergence of the model by looking at the samples from the first and second halves of the run and ensuring that they both follow the same distribution.


\section{Results}
\label{sect:results}

\begin{figure}[h!]
\begin{center}
\includegraphics[height=7.3in]{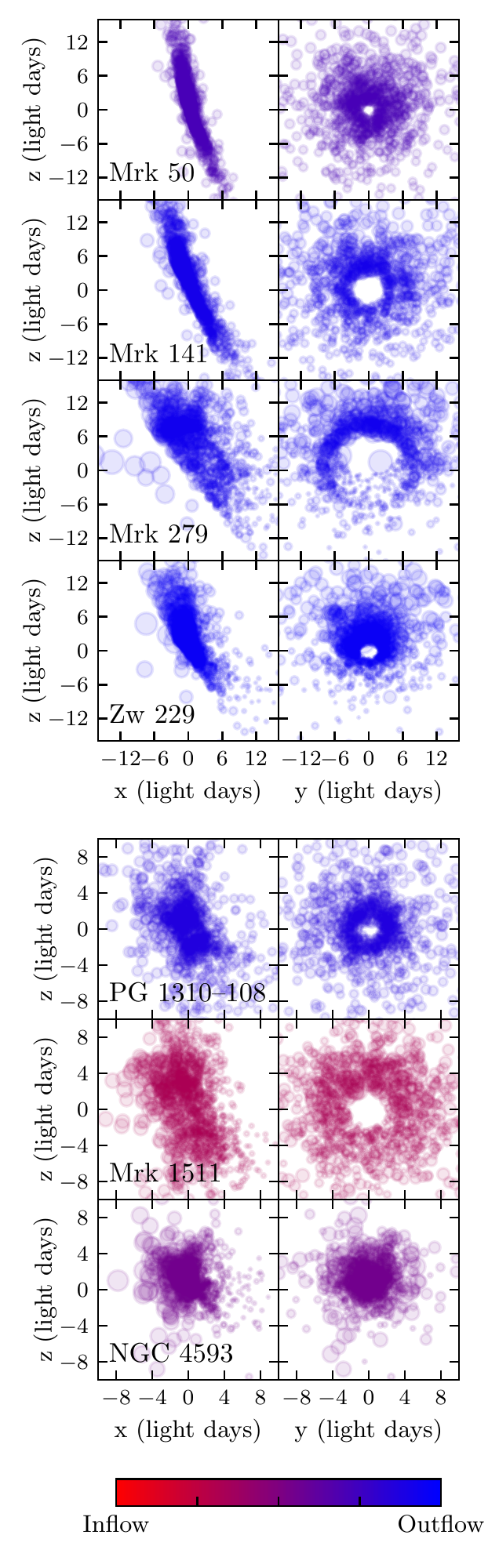}
\caption{Geometries of the \Hb-emitting BLR for each object, drawn from the posterior samples.
In each panel, the observer is viewing the BLR from the positive $x$-axis.
Each circle corresponds to one point particle in the model, and the size of the circle corresponds to the amount of line emission coming from that particle.
The left panels show an edge-on view, while the right panels show a face-on view.
The geometries are color-coded to indicate whether the BLR dynamics exhibit inflow or outflow.
\label{fig:geos}}
\end{center}
\end{figure}

\begin{figure*}[h!]
\begin{center}
\includegraphics[width=0.9\textwidth]{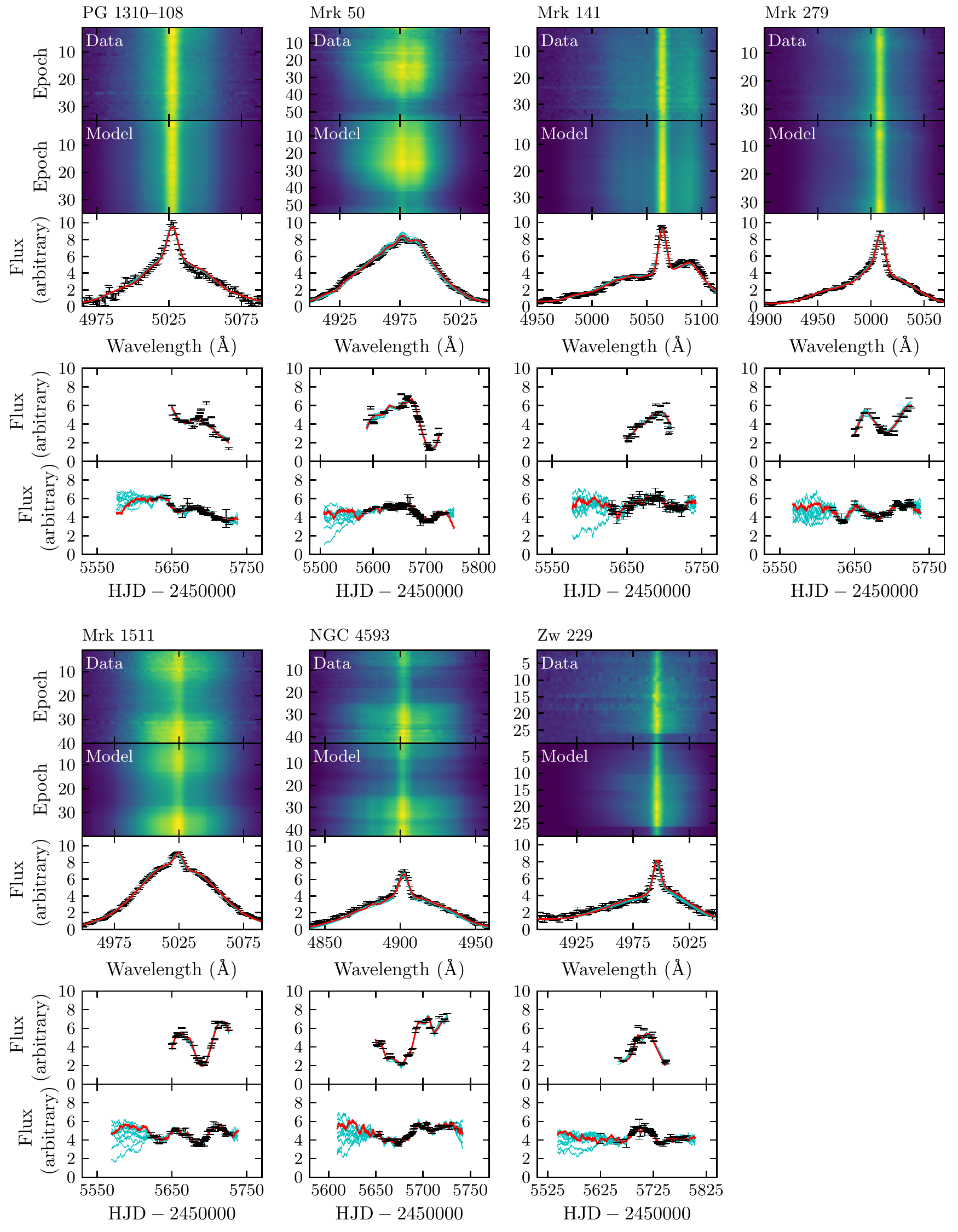}
\caption{Model fits to the \Hb\ line profile, integrated \Hb\ flux, and AGN continuum flux.
From left to right, the panels show models for \iisa, \mrka, \mrkb, \mrkc, \mrkd, \ngca, and \zwia.
Within each panel, numbered 1--5 from top to bottom, we have the following.
\textit{Panels 1 and 2}: The observed \Hb\ emission-line profile by observation epoch and the profile produced by one sample of the BLR and continuum model.
\textit{Panel 3}: The observed \Hb\ profile of one randomly chosen epoch (black), and the corresponding profile (red) produced by the model in Panel 2.
The cyan lines show the \Hb\ profile produced by three other randomly chosen models.
\textit{Panels 4 and 5}: Time series of the observed integrated \Hb\ and continuum flux (black), and the model fits to these light curves (red), corresponding to the model shown in Panel 2.
The cyan lines show five other model examples.
\label{fig:display_all}}
\end{center}
\end{figure*}

\begin{figure*}[h!]
\begin{center}
\includegraphics[width=1\textwidth]{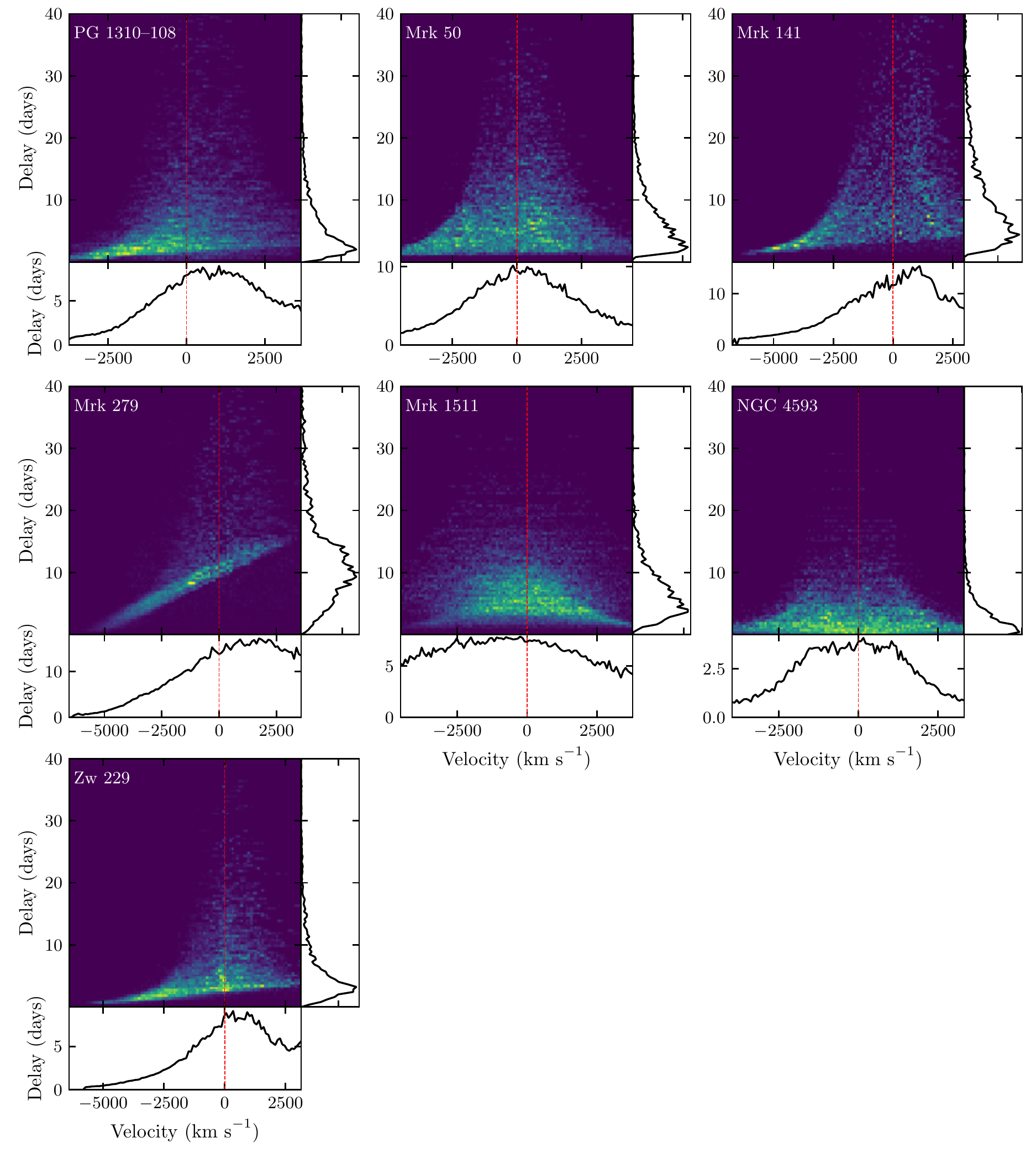}
\caption{Velocity-resolved transfer functions for each AGN, drawn from the posterior samples and selected to be representative of the full posterior samples.
In the right-hand panels, we show the velocity-integrated transfer function, and the bottom panel shows the average time lag for each velocity pixel.
\label{transfer_all}}
\end{center}
\end{figure*}

In this section, we discuss the dynamical modeling results for our sample of seven LAMP 2011 AGNs.
For each AGN, we show a geometric model of the BLR from the posterior sample, chosen to be typical of the geometries in the full posterior sample (Figure \ref{fig:geos}).
We also show randomly chosen model fits to the AGN \Hb\ profile, the integrated \Hb\ flux light curve, and the continuum light curve in Figure \ref{fig:display_all}.
In Figure \ref{transfer_all}, we show velocity-resolved transfer functions for each AGN, created using the same model as in Figure \ref{fig:geos}.

\begin{figure*}[h!]
\begin{center}
\includegraphics[width=\posteriorwidth]{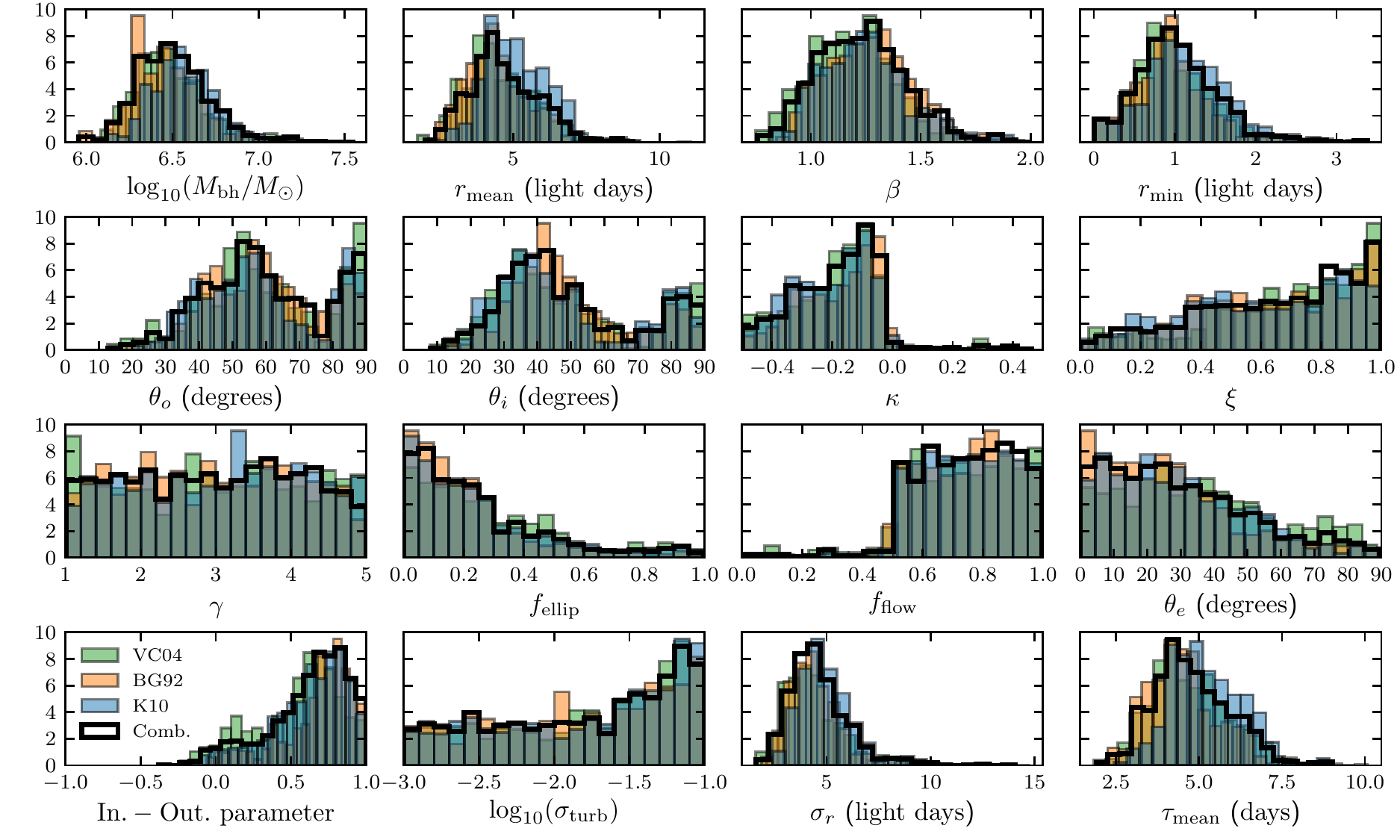}  
\caption{Posterior distributions of the parameters for \iisa.
The blue, orange, and green posterior histograms are for the runs using the spectral decompositions using the \kov, \bg, and \veron\ \feii\ templates, respectively.
The black line shows the combined posterior PDF.
\label{fig_post_iisz10}}
\end{center}
\end{figure*}

\begin{figure*}[h!]
\begin{center}
\includegraphics[width=\posteriorwidth]{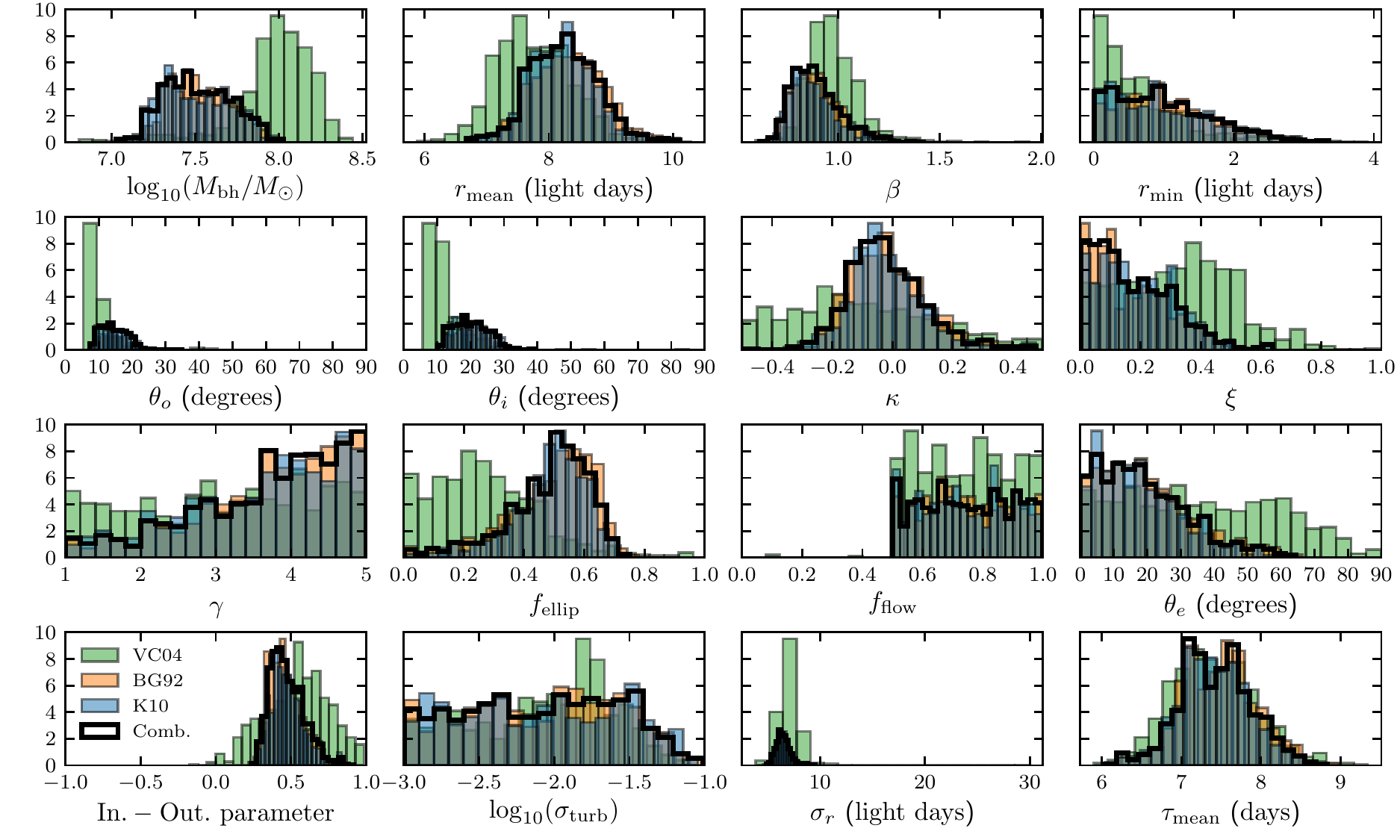}  
\caption{Posterior distributions of the parameters for Mrk 50. For this AGN, the combined posterior was created using only the \kov\ and \bg\ \feii templates.
\label{fig_post_mrk50}}
\end{center}
\end{figure*}

\begin{figure*}[h!]
\begin{center}
\includegraphics[width=\posteriorwidth]{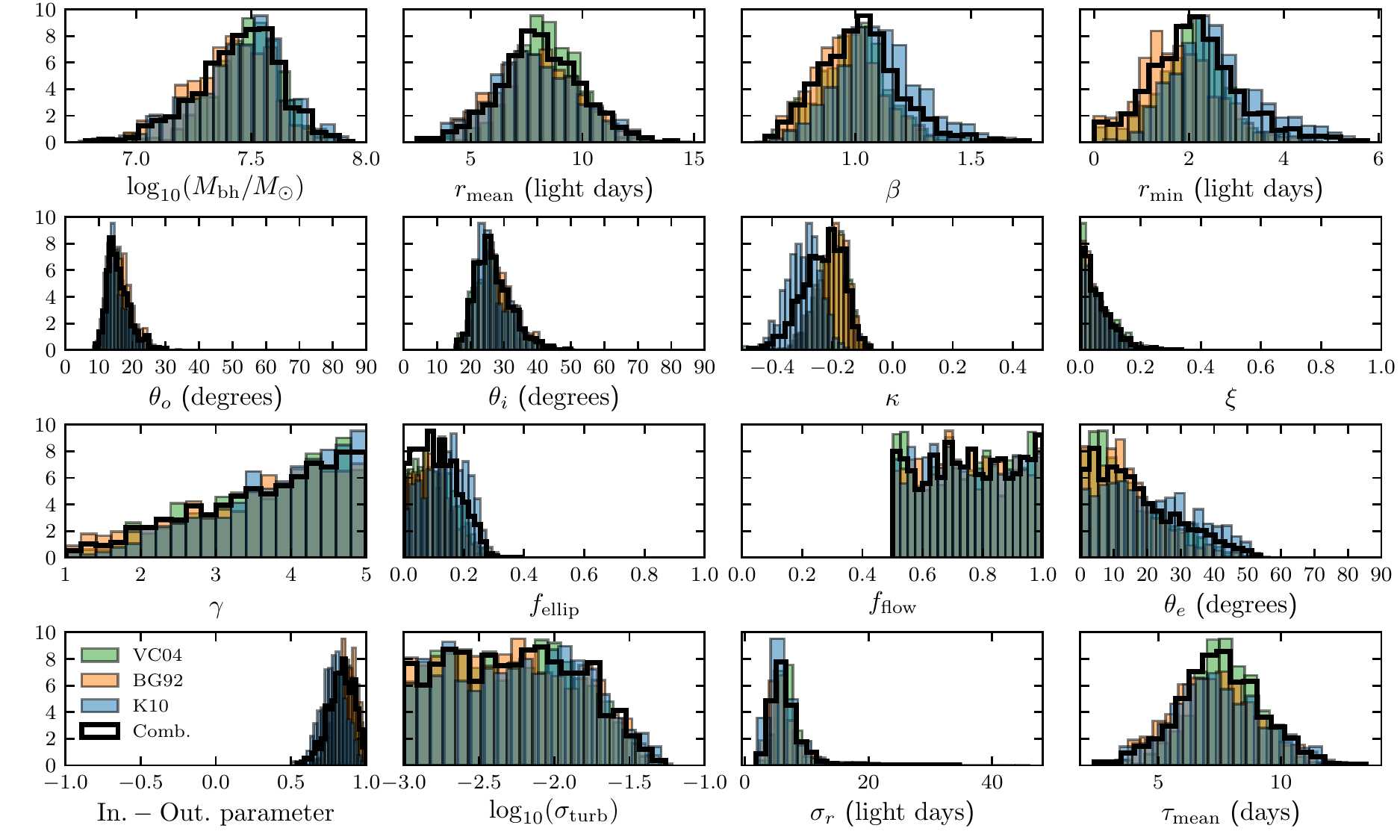}  
\caption{Posterior distributions of the parameters for Mrk 141.
\label{fig_post_mrk141}}
\end{center}
\end{figure*}

\begin{figure*}[h!]
\begin{center}
\includegraphics[width=\posteriorwidth]{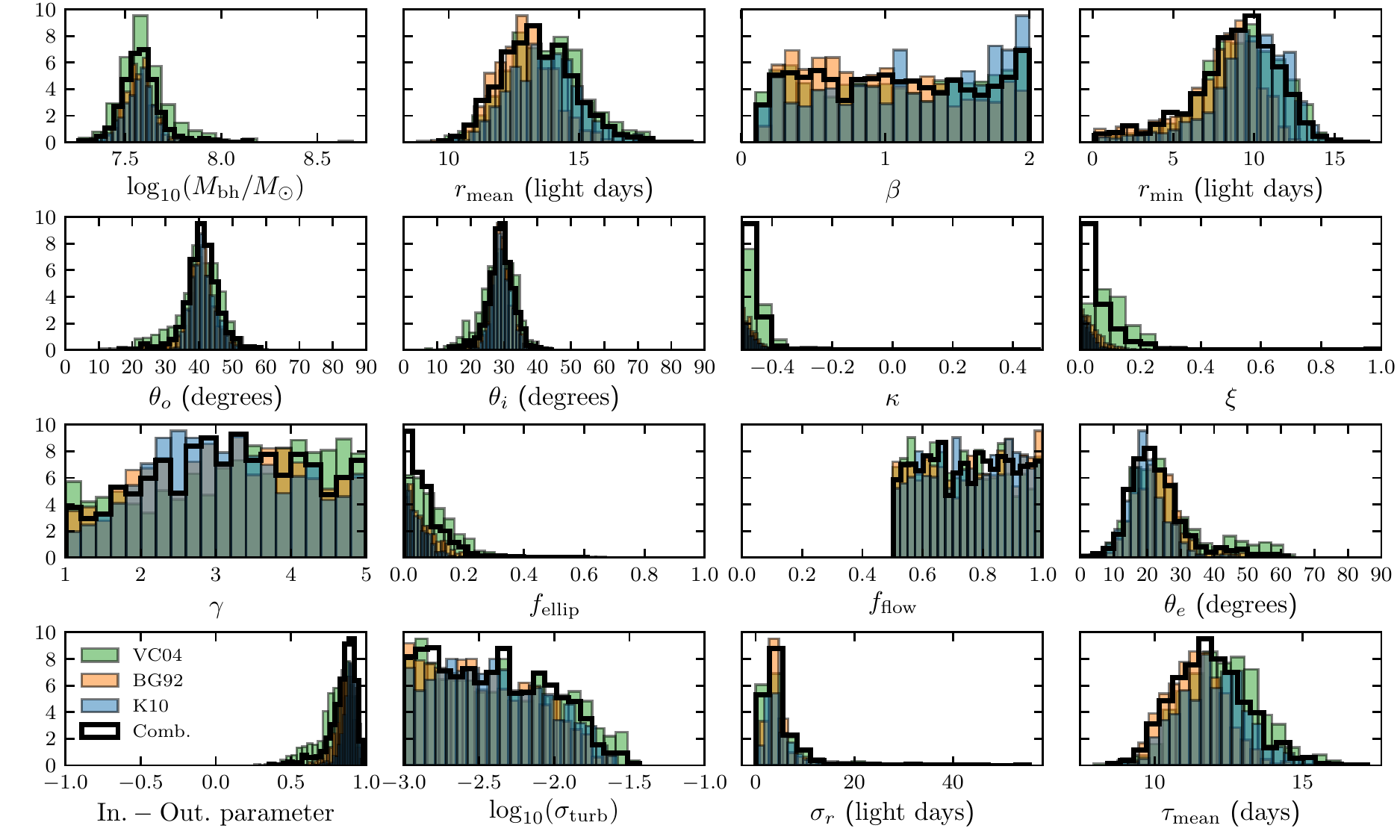}  
\caption{Posterior distributions of the parameters for Mrk 279.
\label{fig_post_mrk279}}
\end{center}
\end{figure*}

\begin{figure*}[h!]
\begin{center}
\includegraphics[width=\posteriorwidth]{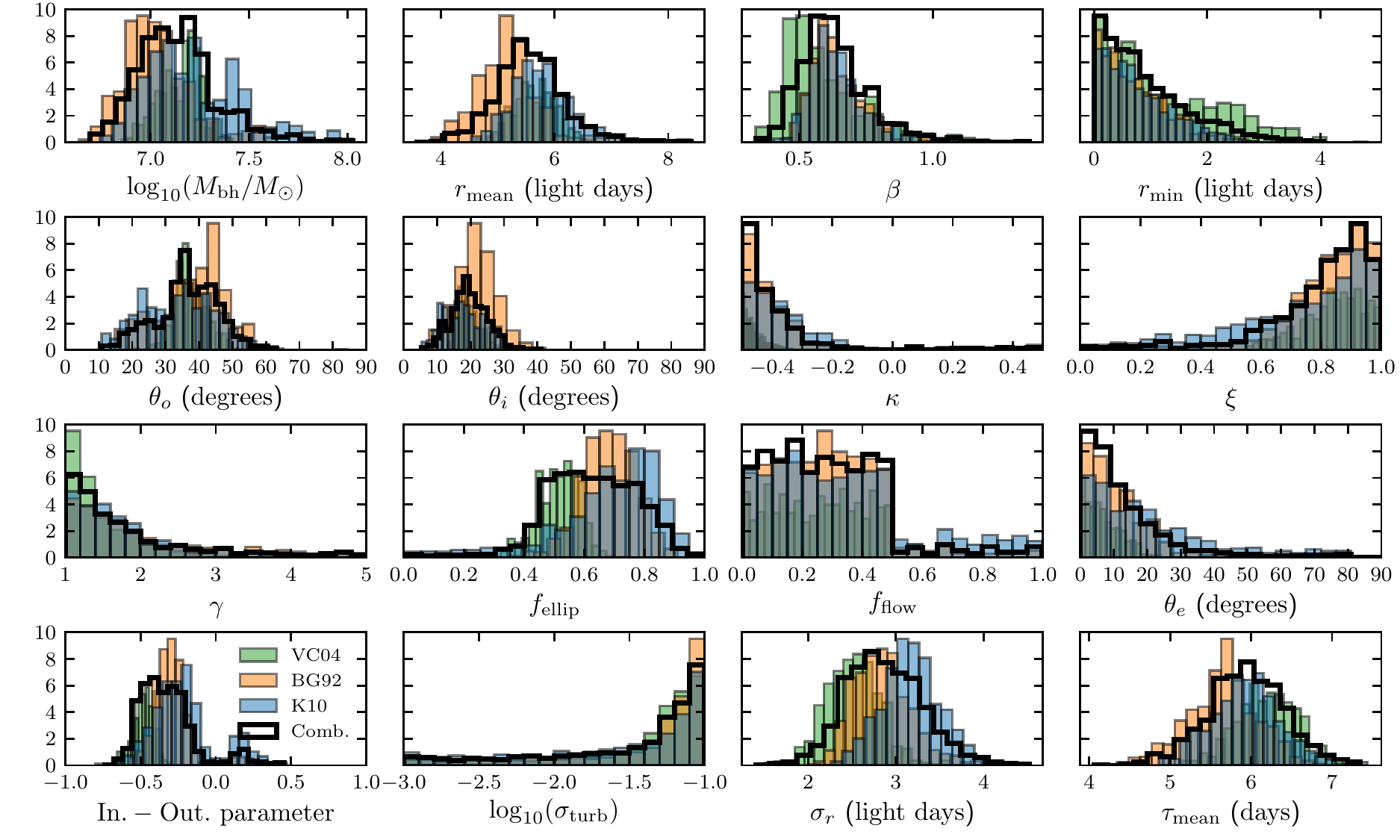}  
\caption{Posterior distributions of the parameters for Mrk 1511.
\label{fig_post_mrk1511}}
\end{center}
\end{figure*}

\begin{figure*}[h!]
\begin{center}
\includegraphics[width=\posteriorwidth]{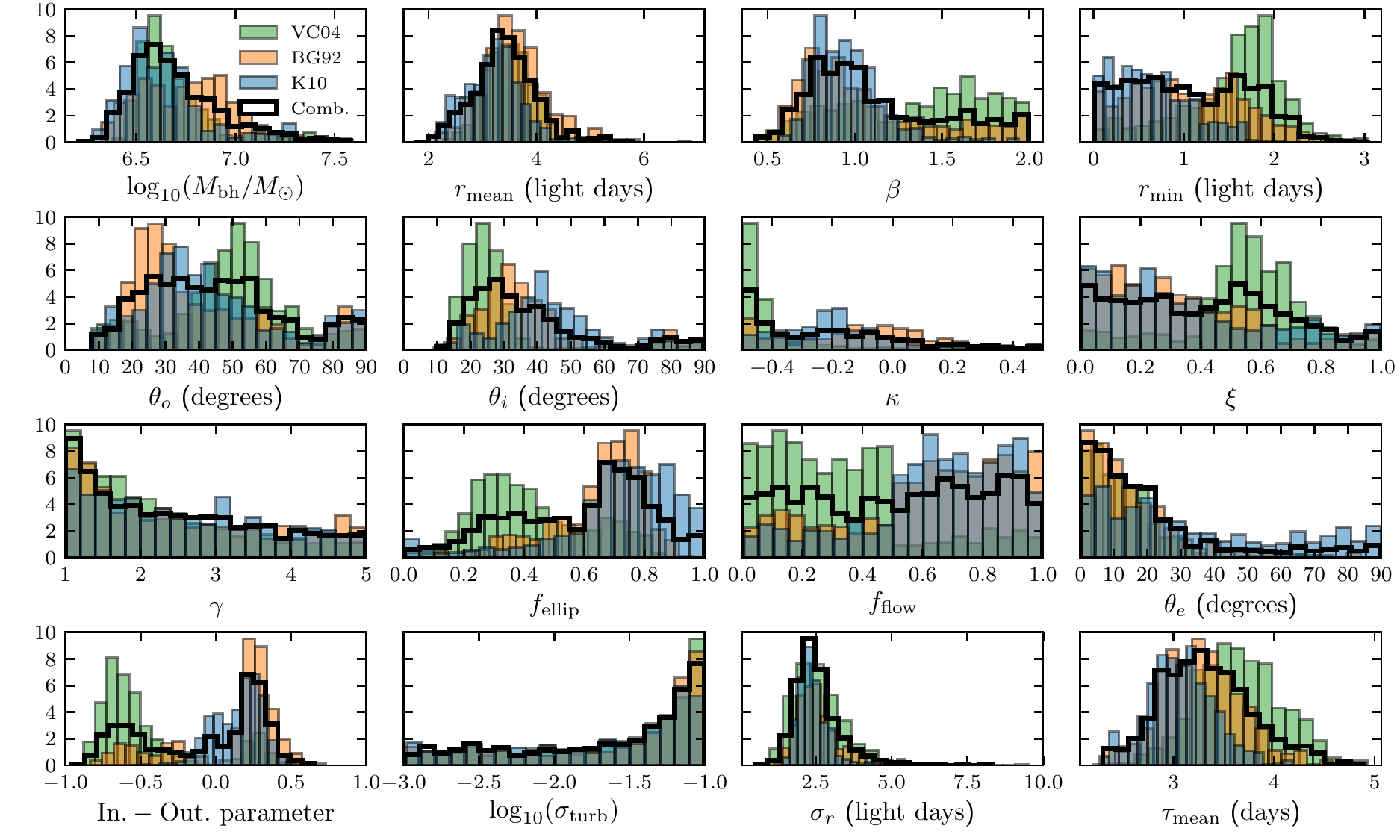}  
\caption{Posterior distributions of the parameters for NGC 4593.
\label{fig_post_ngc4593}}
\end{center}
\end{figure*}

\begin{figure*}[h!]
\begin{center}
\includegraphics[width=\posteriorwidth]{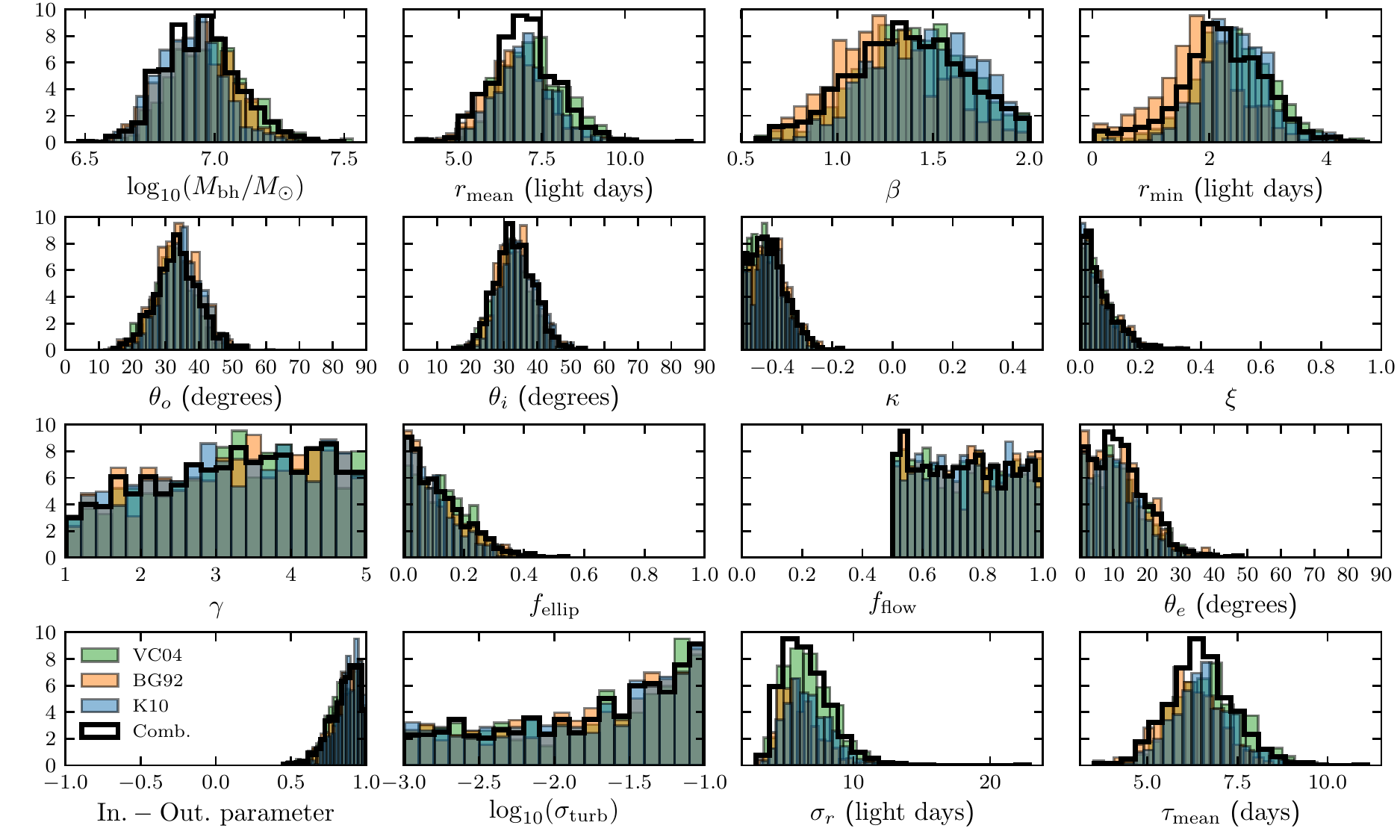}  
\caption{Posterior distributions of the parameters for \zwia.
\label{fig_post_zw229}}
\end{center}
\end{figure*}

In Figures \ref{fig_post_iisz10}--\ref{fig_post_zw229}, we give the posterior distributions of the key model parameters.
We also include a parameter to summarize whether the overall dynamics indicate inflowing or outflowing gas, defined such that $1$ and $-1$ are purely radial outflow and inflow, respectively:
\begin{align}
{\rm In.-Out.} = {\rm sgn}(f_{\rm flow} - 0.5) \times (1 - f_{\rm ellip}) \times \cos(\theta_e),
\end{align}
where ${\rm sgn}$ is the sign function.
The median values and 68\% confidence intervals for all parameter are summarized in Table \ref{table_results}.
When the posterior PDFs are one-sided, we give upper and lower 68\% confidence limits.

\subsection{\iisa}

The top two panels of Figure \ref{fig:display_all} show how the \Hb\ line profile changed over the course of the observing campaign for both the data and one of the models drawn randomly from the posterior sample.
The model fits the data very well in the core of the line, but misses some of the detailed structure in the blue wing.
This is also visible in the third panel where we show a model fit to the \Hb\ spectrum for one epoch.
In the fourth panel, we show the integrated \Hb\ flux over the course of the campaign.
The \Hb\ light curve for \iisa\ is relatively short, owing to the loss of many nights to poor weather, but there is a clear variability signal.
The models were able to fit the overall shape of the \Hb\ light curve, but the details are poorly modeled.

The posterior distributions in Figure \ref{fig_post_iisz10} give a radial distribution of the \iisa\ \Hb\ emission that is steeper than exponential with a shape parameter $\beta = $ \iisabeta\ for the Gamma distribution.
The distribution is shifted from the origin by a minimum radius $r_{\rm min} = $ \iisarmin\ light days and has a mean radius $r_{\rm mean} = $ \iisarmean\ light days and a radial width $\sigma_r = $ \iisasigmar\ light days.
The mean lag is $\tau_{\rm mean} = $ \iisataumean\ days, which is smaller than the cross-correlation measurement $\tau_{\rm cen} = 7.20^{+2.41}_{-3.11}$ days from Barth et al. (2018, in preparation), but is consistent to within the uncertainties.
We note that the uncertainties on both $\tau_{\rm mean}$ and $\tau_{\rm cen}$ are relatively high.
It is possible that the true \Hb\ lag is longer than measured, and that the short spectroscopic monitoring campaign biases our results toward shorter lags.
The distributions for the opening and inclination angles exhibit multiple solutions.
The \Hb-emitting region is inferred to be either a thick disk with opening angle near $50^\circ$ or spherical with opening angle approaching $90^\circ$.
An example of the former is shown in Figure \ref{fig:geos}.

There is no preference for emission to be concentrated near the faces of the disk ($\gamma = $ \iisagamma).
Despite the median and 68\% confidence interval suggesting only a slight preference for emission from the far side of the BLR ($\kappa = $ \iisakappa), almost none of the posterior samples have $\kappa > 0$, ruling out the possibility of preferential emission from the near side.
Solutions with a transparent BLR midplane are preferred slightly over those with an opaque midplane ($\xi = $ \iisaxi).

\newcommand{\resultstablecomment}{Median values and 68\% confidence intervals for the main BLR geometry and dynamics model parameters.
Upper and lower 68\% confidence limits are given when the posterior PDF is one-sided.
Note that $r_{\rm out}$ is a fixed parameter, so we do not include uncertainties.}
\begin{deluxetable*}{lccccccc}
\tablecaption{BLR Model Parameter Values}
\tablehead{ 
\colhead{Parameter} &
\colhead{\mrka} &
\colhead{\mrkb} &
\colhead{\mrkc} &
\colhead{\mrkd} &
\colhead{\ngca} &
\colhead{\iisa} &
\colhead{\zwia}
}
\startdata
$r_{\rm out}$ (light days)      & \mrkarmax           & \mrkbrmax           & \mrkcrmax           & \mrkdrmax           & \ngcarmax           & \iisarmax           & \zwiarmax           \\
$r_{\rm mean}$ (light days)     & \mrkarmean          & \mrkbrmean          & \mrkcrmean          & \mrkdrmean          & \ngcarmean          & \iisarmean          & \zwiarmean          \\
$r_{\rm median}$ (light days)   & \mrkarmedian        & \mrkbrmedian        & \mrkcrmedian        & \mrkdrmedian        & \ngcarmedian        & \iisarmedian        & \zwiarmedian        \\
$r_{\rm min}$ (light days)      & \mrkarmin           & \mrkbrmin           & \mrkcrmin           & \mrkdrmin           & \ngcarmin           & \iisarmin           & \zwiarmin           \\
$\sigma_{r}$ (light days)       & \mrkasigmar         & \mrkbsigmar         & \mrkcsigmar         & \mrkdsigmar         & \ngcasigmar         & \iisasigmar         & \zwiasigmar         \\
$\tau_{\rm mean}$  (days)       & \mrkataumean        & \mrkbtaumean        & \mrkctaumean        & \mrkdtaumean        & \ngcataumean        & \iisataumean        & \zwiataumean        \\
$\tau_{\rm median}$  (days)     & \mrkataumedian      & \mrkbtaumedian      & \mrkctaumedian      & \mrkdtaumedian      & \ngcataumedian      & \iisataumedian      & \zwiataumedian      \\
$\beta$                         & \mrkabeta           & \mrkbbeta           & \mrkcbeta           & \mrkdbeta           & \ngcabeta           & \iisabeta           & \zwiabeta           \\
$\theta_o$ (degrees)            & \mrkathetao         & \mrkbthetao         & \mrkcthetao         & \mrkdthetao         & \ngcathetao         & \iisathetao         & \zwiathetao         \\
$\theta_i$ (degrees)            & \mrkathetai         & \mrkbthetai         & \mrkcthetai         & \mrkdthetai         & \ngcathetai         & \iisathetai         & \zwiathetai         \\
$\kappa$                        & \mrkakappa          & \mrkbkappa          & \mrkckappa          & \mrkdkappa          & \ngcakappa          & \iisakappa          & \zwiakappa          \\
$\gamma$                        & \mrkagamma          & \mrkbgamma          & \mrkcgamma          & \mrkdgamma          & \ngcagamma          & \iisagamma          & \zwiagamma          \\
$\xi$                           & \mrkaxi             & \mrkbxi             & \mrkcxi             & \mrkdxi             & \ngcaxi             & \iisaxi             & \zwiaxi             \\
$\log_{10}(M_{\rm BH}/M_\odot)$ & \mrkalogmbh         & \mrkblogmbh         & \mrkclogmbh         & \mrkdlogmbh         & \ngcalogmbh         & \iisalogmbh         & \zwialogmbh         \\
$f_{\rm ellip}$                 & \mrkafellip         & \mrkbfellip         & \mrkcfellip         & \mrkdfellip         & \ngcafellip         & \iisafellip         & \zwiafellip         \\
$f_{\rm flow}$                  & \mrkafflow          & \mrkbfflow          & \mrkcfflow          & \mrkdfflow          & \ngcafflow          & \iisafflow          & \zwiafflow          \\
$\theta_e$ (degrees)            & \mrkathetae         & \mrkbthetae         & \mrkcthetae         & \mrkdthetae         & \ngcathetae         & \iisathetae         & \zwiathetae         \\
${\rm In.-Out.~param}$          & \mrkainflowoutflow  & \mrkbinflowoutflow  & \mrkcinflowoutflow  & \mrkdinflowoutflow  & \ngcainflowoutflow  & \iisainflowoutflow  & \zwiainflowoutflow  \\
$\sigma_{\rm turb}$             & \mrkasigmaturb      & \mrkbsigmaturb      & \mrkcsigmaturb      & \mrkdsigmaturb      & \ngcasigmaturb      & \iisasigmaturb      & \zwiasigmaturb
\enddata
\tablecomments{\resultstablecomment
\label{table_results}}
\end{deluxetable*} 

Dynamically, \iisa\ is best described by models in which few particles are in near-circular orbits ($f_{\rm ellip} $ \iisafellip).
The remaining particles are on outflowing orbits, given by $f_{\rm flow} = $ \iisafflow, where $f_{\rm flow}$ is a binary parameter with $f_{\rm flow} < 0.5$ indicating inflow and $f_{\rm flow} > 0.5$ indicating outflow.
The full posterior PDF shows almost no solutions with inflow.
The outflowing orbits have velocities drawn from a distribution whose center is rotated $\theta_e =$ \iisathetae\ degrees from the radial escape velocity toward the circular velocity; thus, more than half of the orbits are actually bound.
Finally, there is a small contribution from macroturbulent velocities, with $\sigma_{\rm turb} = $ \iisasigmaturb\ times the circular velocity.

The preference for outflow is visible in the transfer function (Figure \ref{transfer_all}), in which there is a slight upward-angled structure.
This is a signature one would expect for radially outflowing gas (see, e.g.,~\citealt{welsh91}), indicating that the particles with the shortest lags, which are the particles directly between the ionizing source and the observer, are preferentially blueshifted, while those with the longest lags on the far side of the source are preferentially redshifted.

The black hole mass for \iisa\ is found to be $\log_{10}(M_{\rm BH}/M_\odot) = $ \iisalogmbh.
This value was previously measured by \citet{Schulze+10} and is reported in \citet{Busch++14} as $\log_{10}(M_{\rm BH}/M_\odot) = 7.33 \pm 0.3$.
Their measurement was made using the BLR radius estimated from the BLR size-luminosity relationship \citep{bentz09a} combined with the line dispersion and mean scale factor $f_{{\rm mean},{\sigma}} = 3.85\pm 1.15$ from \citet{collin06}.
Our results suggest that the scale factor for this object, when using the line dispersion measured in the mean spectrum, should be \logfmeansigma$ = -0.20^{+0.27}_{-0.21}$ ($f = 0.63^{+0.54}_{-0.24}$), which is much smaller than the values that are typically used \citep[e.g., $f_{{\rm mean},{\sigma}} = 3.85\pm 1.15$;][]{collin06}.
The lower scale factor in this object illustrates the importance of calculating scale factors on an individual AGN basis.
The use of a mean scale factor leads to underestimates and overestimates of $M_{\rm BH}$ in objects with higher and lower intrinsic $f$ values, respectively.

\subsection{Mrk 50}

The BLR in \mrka\ was previously modeled by \citet{pancoast12} using the same LAMP 2011 data and an earlier version of the model used in this paper.
Their model did not include the parameters $\gamma$ or $\xi$ which help introduce asymmetries in the broad-line profile, and the dynamics component did not allow for macroturbulent velocities or the possibility of unbound inflowing or outflowing gas.
We also include an additional narrow-line component in the model and leave the AGN redshift as a free parameter.
The spectral decomposition they use is from \citet{barth11b} and uses the \bg\ \feii\ template.
In our analysis, we use the spectra found adopting both the \kov\ and \bg\ templates, but since both are in very good agreement (Figure \ref{fig:decomps}), we do not expect this to introduce any discrepancies in our measurements.

As discussed in Section \ref{sect:decomp}, the \veron\ \feii\ template produced poor fits for some of the Mrk 50 spectra, so we chose to only use the results from the \kov\ and \bg\ templates for our final analysis.
For completeness, we include the results from the \veron\ template in Figure \ref{fig_post_mrk50}, but the combined posteriors shown by the black lines are computed using only the other two templates.

The model for \mrka\ fits the shape of the \Hb\ emission line very well, as shown in Figure \ref{fig:display_all}.
The large-scale fluctuations in the integrated \Hb\ flux are well captured, but the small peak in flux in the first few epochs is not recovered.

The radial distribution of \Hb\ emission in \mrka\ is between Gaussian and exponential, with a shape parameter $\beta = $ \mrkabeta.
The radial distribution is shifted by a minimum radius $r_{\rm min} = $ \mrkarmin\ light days, and it has a mean radius $r_{\rm mean} = $ \mrkarmean\ light days and a radial width $\sigma_r = $ \mrkasigmar\ light days.
The mean lag is slightly smaller than this with $\tau_{\rm mean} = $ \mrkataumean\ days, which is again smaller than the cross-correlation measurement $\tau_{\rm cen} = 8.66^{+1.63}_{-1.51}$ days from Barth et al. (2018, in preparation), but is consistent to within the uncertainties.
The opening and inclination angles are well constrained and prefer a slightly thick disk geometry, oriented close to face-on ($\theta_o$ = \mrkathetao, $\theta_i$ = \mrkathetai\ degrees).
An example of this geometry is shown in Figure \ref{fig:geos}.
There is a slight preference for the emission to be concentrated near the faces of the disk, with $\gamma = $ \mrkagamma, but uniform emission throughout the disk is not ruled out.
The disk midplane is mostly opaque ($\xi = $ \mrkaxi) and the relative strength of emission from the near and far side of the BLR is equal ($\kappa = $ \mrkakappa).

Our geometric model results are generally in good agreement with those of \citet{pancoast12}.
The largest discrepancy is in the opening and inclination angles, where our results show slightly larger values for both angles.
We do find a disk midplane that is mostly opaque, which was not possible in the earlier version of the model.
The 2D posterior samples for these parameters show that smaller opening and inclination angles are preferred for higher values of $\xi$ (transparanet midplane), so it is possible that this new flexibility is the main cause of the discrepancy.

Dynamically, the model of \mrka\ prefers solutions in which half of the particles are on near-circular elliptical orbits ($f_{\rm ellip} = $ \mrkafellip).
The remaining particles have velocities drawn from a distribution in $v_r - v_\phi$ space with an outflowing radial component ($f_{\rm flow} = $ \mrkafflow).
The center of this distribution is rotated $\theta_e = $ \mrkathetae\ degrees from the radially outflowing escape velocity toward the circular velocity.
The contribution of macroturbulent velocities is minimal, with $\sigma_{\rm turb} = $ \mrkasigmaturb.

We find the black hole mass to be $\log_{10}(M_{\rm BH}/M_\odot) = $ \mrkalogmbh.
This is consistent with the \citet{pancoast12} measurement of $7.57^{+0.44}_{-0.27}$, despite the different models.
This is reassuring but also perhaps not surprising, given that we do not find a significant inflow or outflow component that our model would be able to better describe than the previous version.

\subsection{Mrk 141}
\label{sect:mrk141}

The dataset for \mrkb\ is of relatively low quality owing to many spectroscopic observing nights lost to poor weather.
The integrated \Hb\ light curve is relatively short and there are not many strong variability features, but there is one large increase and decrease in flux over the course of the campaign.
The models are able to fit this overall feature, but do not fit the smaller fluctuations on scales of a few days.
The models are able to fit the shape of the \Hb\ profile very well.

The posterior PDFs from runs using all three \feii\ templates agree very well for most parameters, with the largest discrepancy coming from the parameter $\kappa$.
The radial distribution of the BLR in \mrkb\ is roughly exponential, with shape parameter $\beta = $ \mrkbbeta, and is shifted from the origin by $r_{\rm min} = $ \mrkbrmin\ light days.
The mean radius is $r_{\rm mean} = $ \mrkbrmean\ light days, and the radial width of the disribution is $\sigma_r = $ \mrkbsigmar\ light days.
The mean lag is very similar to $c\times r_{\rm mean}$, with $\tau_{\rm mean} = $ \mrkbtaumean\ days.
This value is consistent with the cross-correlation measurement $\tau_{\rm cen} = 5.63^{+8.27}_{-1.65}$ days.

The opening and inclination angles indicate a thick disk ($\theta_o = $ \mrkbthetao\ degrees) inclined $\theta_i = $ \mrkbthetai\ degrees relative to the observer.
As with \mrka, there is a small preference for \Hb\ emission to be concentrated near the faces of the disk, but uniform emission is not ruled out ($\gamma = $ \mrkagamma).
The midplane of the disk is opaque ($\xi $ \mrkbxi).
All three posterior PDFs generated using each \feii\ template indicated a preference for emission from the far side of the BLR, but the result is slightly more pronounced using the \kov\ template results.
In the combined posterior, $\kappa = $ \mrkbkappa.

Dynamically, $< 20$\% of the \Hb-emitting BLR is on near-circular elliptical orbits ($f_{\rm ellip} = $ \mrkbfellip).
The remainder have velocities drawn from a distribution around the outflowing escape velocity, rotated $\theta_e = $ \mrkbthetae\ degrees toward the circular velocity in the $v_r - v_\phi$ plane.
Macroturbulent velocities are not significant in \mrkb\, with $\sigma_{\rm turb} = $ \mrkbsigmaturb.

We find the black hole mass in \mrkb\ to be $\log_{10}(M_{\rm BH}/M_\odot) = $ \mrkblogmbh.
Previous measurements of the black hole mass have been made using the BLR radius-luminosity relation and the FWHM of the \Hb\ line, which find $\log_{10}(M_{\rm BH}/M_\odot) = 7.53$ \citep{Castello-mor++17} and $7.85$ \citep{Li++08}.
These studies do not report uncertainties, but assuming a typical uncertainty of $0.4$ dex arising from the scatter in the $r-L$ relation and the uncertainty in the scale factor used, our result is consistent with both measurements.

\subsection{Mrk 279}

The \Hb\ line profile for \mrkc\ is modeled very well, with only a slight discrepancy at the red side of the core of the line.
The large timescale variations of the integrated \Hb\ line flux are well captured, but the model is unable to reproduce the smaller fluctuations on the order of days.

The \mrkc\ \Hb-emitting region has a radial profile that is poorly determined, with anything from a narrow Gaussian to a steeper than exponential profile being allowed ($\beta = $ \mrkcbeta).
The minimum radius is found to be large, with $r_{\rm min} = $ \mrkcrmin\ light days.
The 2D posterior distributions of these two parameters shows that smaller values of the minimum radius ($<5$ light days) are allowed when $\beta < 0.5$ (narrow Gaussian), but for wider Gaussian and steep exponential profiles, the minimum radius is robustly determined.

The opening angle and inclination angle for \mrkc\ are $\theta_o = $ \mrkcthetao\ degrees and $\theta_i = $ \mrkcthetai\ degrees, respectively, indicating a thick disk that is slightly inclined relative to the observer.
Based on the full posterior PDF for $\gamma$, there is no preference for emission to either be concentrated near the faces of the disk or for it to be uniform throughout.
There is a very strong preference for emission from the far side of the BLR ($\kappa $ \mrkckappa), and the midplane of the disk is fully opaque ($\xi $ \mrkcxi).
This shows up clearly in the geometric model (Figure \ref{fig:geos}) in that there are very few points visible on the bottom-left half of the edge-on view, and the points that are farther from the observer are larger, representing the relative strength of the emission.

In \mrkc, models with almost no particles on near-circular orbits are preferred ($f_{\rm ellip} $ \mrkcfellip).
Instead, velocities are drawn from a distribution rotated $\theta_e = $ \mrkcthetae\ degrees from the radially outflowing escape velocity toward the circular velocity.
The upward angled outflow signature discussed in Section \ref{sect:mrk141} is prominent in the \mrkc\ transfer fuction (Figure \ref{transfer_all}) and is the strongest for the whole LAMP 2011 sample.
There is no indication of a significant contribution from macroturbulent velocities ($\sigma_{\rm turb} = $ \mrkcsigmaturb).

We find a black hole mass of $\log_{10}(M_{\rm BH}/M_\odot) = $ \mrkclogmbh\ in \mrkc.
This object was spectroscopically monitored in 1988, when \citet{Mao++90} found the size of the \Hb-emitting BLR to be $11 \pm 3$ light days.
\citet{San++01} observed \mrkc\ from 1996 to 1997 and measured an \Hb\ time lag of $\tau = 16^{+5.3}_{-5.6}$ days.
\citet{Pet++04} later re-analyzed both datasets and, assuming a value of 5.5 for $f_{{\rm rms},{\sigma}}$, measured the black hole mass to be $M_\text{BH}/(10^6\,M_\odot) = 34.9\pm 9.2$ [$\log_{10}(M_{\rm BH}/M_\odot) = 7.54^{+0.10}_{-0.13}$].
Our results for the size, time lag, and black hole mass are consistent with all of these results.
We note that the true BLR size and \Hb\ time lag may have changed between our three campaigns owing to changes in the AGN continuum, but the black hole mass should remain the same.
\mrkc\ was also analyzed by \citet{Castello-mor++17} using the radius-luminosity relationship as a BLR size estimator, and they found $\log_{10}(M_{\rm BH}/M_\odot) = 7.97$.
As in the case of \mrkb, our results are consistent with this measurement when realistic uncertainties are assumed for the previous estimates.

\subsection{Mrk 1511}

Both the \Hb\ emission-line shape and the variability in the integrated \Hb\ flux are very well fit by the models for \mrkd.
However, the posterior PDFs in Figure \ref{fig_post_mrk1511} indicate that the \Hb\ spectra produced using the three \feii\ templates give somewhat different modeling results.
The right panels of Figure \ref{fig:decomps} show that there is a discrepancy between the three \Hb\ profiles that is on the order of the size of the flux uncertainty.
This is likely a result of the strong \feii\ contribution to the AGN spectrum, shown by the green line in the left panel.
Since the \kov\ template is made up of five individual components whose strengths are given by five free parameters, it has more flexibility to fit asymmetries in the \feii\ emission than the other templates.
This can result in asymmetries in the resulting \Hb\ spectrum.
Since the model results are still consistent with each other, we choose to combine them, the result of which is larger parameter uncertainties.

The \Hb-emitting BLR in \mrkd\ has a radial profile that is between Gaussian and exponential ($\beta = $ \mrkdbeta) and is shifted from the central ionizing source by $r_{\rm min} = $ \mrkdrmin\ light days.
The mean radius is $r_{\rm mean} = $ \mrkdrmean\ light days, and the radial thickness is $\sigma_r = $ \mrkdsigmar\ light days.
The mean time lag is $\tau_{\rm mean} = $ \mrkdtaumean\ days, consistent with the cross-correlation lag measurement of $\tau_{\rm cen} = 5.44^{+0.74}_{-0.67}$ days.
The structure of the BLR is best described by a thick disk with opening angle $\theta_o = $ \mrkdthetao\ degrees that is inclined by $\theta_i = $ \mrkdthetai degrees relative to the observer.
The \Hb\ emission is mostly uniform throughout the disk, as opposed to being concentrated near the faces ($\gamma $ \mrkdgamma).
There is a strong preference for emission from the far side of the BLR ($\kappa $ \mrkdkappa), and the disk midplane is transparent ($\xi = $ \mrkdxi).

The largest discrepancy in the results from the three runs comes from the dynamical component of the model.
The results from using the \veron\ \feii\ template indicates that half of the BLR is in near-circular orbits, while the other half is in near-radial infall.
The results from using the \kov\ and \bg\ templates suggest that closer to 3/4 of the BLR is in near-circular orbits.
The remaining 1/4 is in close-to-radial infall, though radial outflow is not fully ruled out.
Combining all three posteriors, we find $f_{\rm ellip} = $ \mrkdfellip\ of the particles on near-circular orbits, with the remaining particles close to radial infall ($f_{\rm flow} = $ \mrkdfflow, $\theta_e = $ \mrkdthetae\ degrees).
Of all the objects in the LAMP 2011 sample, macroturbulent velocities have the highest effect on the BLR dynamics in \mrkd, with $\sigma_{\rm turb} $ \mrkdsigmaturb.
The posterior PDF shows that this value is approaching its prior bound of 0.1, so the contribution may actually be higher.
We find the black hole mass in \mrkd\ to be $\log_{10}(M_{\rm BH}/M_\odot) = $ \mrkdlogmbh.

\subsection{NGC 4593}

The models for \ngca\ fit the observed \Hb\ emission-line shape very well.
Additionally, the models are able to recover almost all of the variation in the \Hb\ light curve, including the short-timescale variations.
Like with \mrkd, the results from using different \feii\ templates show disagreement in some parameters.
Looking at the spectral decomposition for \ngca, there is a significant difference between the \Hb\ profiles produced using the three templates, especially in the wings of the line, owing to the strong \feii\ emission in this object.

The results from using the \kov\ and \bg\ \feii\ templates show that the radial profile of the \ngca\ BLR is between Gaussian and exponential, with the minimum radius poorly constrained.
When the \veron\ template is used, the Gamma function shape parameter is poorly constrained, but the minimum radius is found to be about 1.8 light days.
Combining the posteriors, we find a radial profile that is near exponential ($\beta = $ \ngcabeta) and is shifted from the central ionizing source by $r_{\rm min} = $ \ngcarmin\ light days.
The mean radius is $r_{\rm mean} = $ \ngcarmean\ light days, and the radial thickness is $\sigma_r = $ \ngcasigmar\ light days.
The mean time lag is $\tau_{\rm mean} = $ \ngcataumean\ days, which is consistent with the cross-correlation lag measurement of $\tau_{\rm cen} = 3.54^{+0.76}_{-0.82}$ days.
The structure is best described by a thick disk with opening angle $\theta_o = $ \ngcathetao\ degrees that is inclined by $\theta_i = $ \ngcathetai\ degrees relative to the observer.
The \Hb\ emission is mostly uniform throughout the disk ($\gamma $ \ngcagamma), there is a preference for emission from the far side of the BLR ($\kappa = $ \ngcakappa), and the disk midplane is between transparent and opaque ($\xi = $ \ngcaxi).

Dynamically, the results from the three runs show disagreement in the amounts of inflowing or outflowing gas.
The results from using the \kov\ and \bg\ \feii\ templates are in agreement, with 3/4 of the particles on near-circular orbits, and the remainder on radially outflowing trajectories.
Using the \veron\ template, only 1/4 of the orbits are inferred to be near-circular, with the remainder in near-radial inflow.
Outflowing trajectories, however, are not fully ruled out.
All three runs give the same result that macroturbulent velocities may be important for the dynamics of \ngca\ ($\sigma_{\rm turb} $ \ngcasigmaturb).
As with \mrkd, this parameter is approaching its prior bound of 0.1, meaning the contribution could be larger.

Despite the differences in inferred BLR structure and dynamics, all three runs converge to the same black hole mass of $\log_{10}(M_{\rm BH}/M_\odot) = $ \ngcalogmbh.

\subsection{\zwia}

The spectroscopic monitoring for \zwia\ also suffered from losses owing to poor weather, having the fewest spectroscopic observations in the sample.
The large-scale variability features are driven by the final three data points, with some shorter timescale variability around the peak.
The models are able to recover the variability on both long and short timescales.
The \Hb\ line profile's asymmetric shape is also very well modeled.

The radial profile of the BLR in \zwia\ is steeper than exponential ($\beta = $ \zwiabeta) and is shifted from the central ionizing source by $r_{\rm min} = $ \zwiarmin\ light days.
The mean radius is $r_{\rm mean} = $ \zwiarmean\ light days and the radial thickness is of similar size, $\sigma_r = $ \zwiasigmar\ light days.
The mean lag is inferred to be $\tau_{\rm mean} = $ \zwiataumean\ days, which is consistent with the cross-correlation measurement of $\tau_{\rm cen} = 5.90^{+7.61}_{-2.40}$ days.
\zwia\ was also monitored at Lick Observatory from 2010 June to December, and \citet{barth11} measure an \Hb\ lag of $\tau_{\rm cen} = 3.86^{+0.69}_{-0.90}$ days using these data.
The structure of the BLR is well constrained to be a thick disk ($\theta_o = $ \zwiathetao\ degrees), inclined $\theta_i = $ \zwiathetai\ degrees relative to the observer.
There is little preference for the emission to be either concentrated near the faces of the disk or distributed uniformly throughout the disk.
The emission comes mostly from the far side of the BLR ($\kappa = $ \zwiakappa), and the midplane of the disk is fully opaque ($\xi $ \zwiaxi).

Dynamically, models in which almost all particles are on outflowing trajectories are preferred.
The fraction of the BLR that is on near-circular orbits is $f_{\rm ellip} $ \zwiafellip.
The remaining particles have velocities drawn from a distribution whose center is rotated $\theta_e = $ \zwiathetae\ degrees from the radially outflowing escape velocity toward the circular velocity in the $v_r - v_\phi$ plane.
\citet{barth11} split the \Hb\ profile of \zwia\ into six bins to make velocity-resolved reverberation mapping measurements.
Qualitatively, they find results that are consistent with Keplerian motions, although they do not state conclusive evidence owing to the large error bars on the measurements.
It is inconclusive from the results whether macroturbulent velocities contribute significantly to the BLR dynamics ($\sigma_{\rm turb} = $ \zwiasigmaturb).
This parameter is approaching its prior bound of 0.1, so it is possible that the contribution is higher.

We find the black hole mass in \zwia\ to be $\log_{10}(M_{\rm BH}/M_\odot) = $ \zwialogmbh.
This is consistent with the value of $\log_{10}(M_{\rm BH}/M_\odot) = 7.00^{+0.08}_{-0.12}$ from \citet{barth11}, even though the measured lag changed between the two campaigns.
This is reassuring since, while the BLR size (and hence lag) may change on these timescales due to changes in the ionizing continuum, the black hole mass should remain the same.
The consistency in measurements across the two epochs serves as a test of the two methods.


\section{Discussion}
\label{sect:discussion}

\subsection{Overall Properties of the LAMP 2011 BLRs}

The results from this analysis paint a fairly uniform picture for the geometric structure of the \Hb-emitting BLR in the AGN in this sample.
We note, however, that this sample spans a limited range in luminosity and black hole mass, so BLR geometries may differ for objects outside of this parameter space.
We find the BLR in all objects to be thick disks that are viewed close to face-on.
The radial distribution of particles is typically between Gaussian and exponential, when well determined.
There is a preference for emission from the far side of the BLR in all objects, except for \mrka, whose results allow for the possibility of preferential emission from the near side.
Emission from the far side is what one expects based on photoionization model predictions that \Hb\ is mostly re-emitted back towards the ionizing source \citep{Ferland++92,OBrien++94}.
Most objects allow for the possibility of either uniform emission throughout the disk or concentrated emission near the faces of the disk.
\mrkd, the only object in which this parameter is well constrained, shows a strong preference for uniform emission.
Finally, the parameter determining the transparency of the disk midplane varies across our sample.

Dynamically, we find more variety in our sample.
The contribution of macroturbulent velocities is unconstrained or negligible in most objects, but \mrkd, \ngca, and \zwia\ show a possible significant contribution with $\sigma_{\rm turb}$ approaching its upper prior bound of 0.1.
Three of the objects (\mrka, \mrkd, and \ngca) are found to have more than half of the particles on near-circular orbits, and the rest have almost all particles on either inflowing or outflowing trajectories.
For every object except \mrkd, the particles that are not on near-circular orbits are on outflowing trajectories.
This is in contrast to the results from \citetalias{pancoast14b} and \citetalias{Grier++17}, both of which find mostly near-circular orbits, inflowing orbits, or a combination of the two.
To our knowledge, outflow in the \Hb-emitting BLR has only been observed in three other cases: NGC 3227 \citep{Denney++09}, Mrk 142 \citep{Du++16}, and MCG $+$06$-$26$-$012 \citep{Du++16}.
The model used in this paper is not able to constrain the detailed dynamics of the outflow.
If the outflow is launched and then the gas is left to move under the influence of the black hole's gravity, then the assumption that the black hole's gravity dominates BLR motions holds.
However, significant outward forces due to (for example) radiation pressure might lead us to underestimate the mass \citep{marconi08,marconi09,netzer09,netzer10}.
A more detailed analysis of the dynamics will be necessary to fully understand how outflows affect black hole mass measurements.

\subsection{Scale Factor $f$}
\label{sect:scalefactor}

The scale factor $f$ gives the relationship between the virial product ($M_{\rm vir} = c\tau v^{2}/G$) and the black hole mass ($M_{\rm BH} = f\,M_{\rm vir}$).
This value depends on the structure and physical orientation of the BLR as well as the dynamics and emitting properties of the gas that can affect the asymmetry of the broad-line profile.
We do not expect all of these properties to be the same for every AGN, so each AGN has its own conversion factor between the virial product and the black hole mass.
However, since traditional reverberation mapping cannot recover individual scale factors, an average value is typically used.
This is measured by finding the value that puts AGN black hole mass measurements in alignment with the $M_{\rm BH}-\sigma_*$ relation for quiescent galaxies.
Results from these analyses range from $f_{{\rm rms},{\sigma}} = 2.8^{+0.7}_{-0.5}$ [i.e., $\log_{10}(f_{{\rm rms},{\sigma}}) = 0.44^{+0.10}_{-0.09}$] \citep{graham11} to $f_{{\rm rms},{\sigma}} = 5.9^{+2.1}_{-1.5}$ [i.e., $\log_{10}(f_{{\rm rms},{\sigma}}) = 0.77 \pm 0.13$] \citep{woo13}, though these measurements are generally consistent to within the uncertainties.

Using the direct modeling approach, we can calculate the scale factor $f$ for individual AGNs by combining the black hole masses inferred by our model with line width and BLR size measurements.
In the following sections, we use BLR size measurements based on $r = c\tau$, where the time delay is the cross-correlation measurement $\tau_{\rm cen}$ from Barth et al. (2018, in preparation).
The line widths consist of three measurements: the FWHM measured in the mean spectrum, the line dispersion measured in the mean spectrum, and the line dispersion measured in the rms spectrum.
All line widths for the individual sources presented here come from \citet{barth15}.

\begin{figure}[h!]
\begin{center}
\includegraphics[width=3.1in]{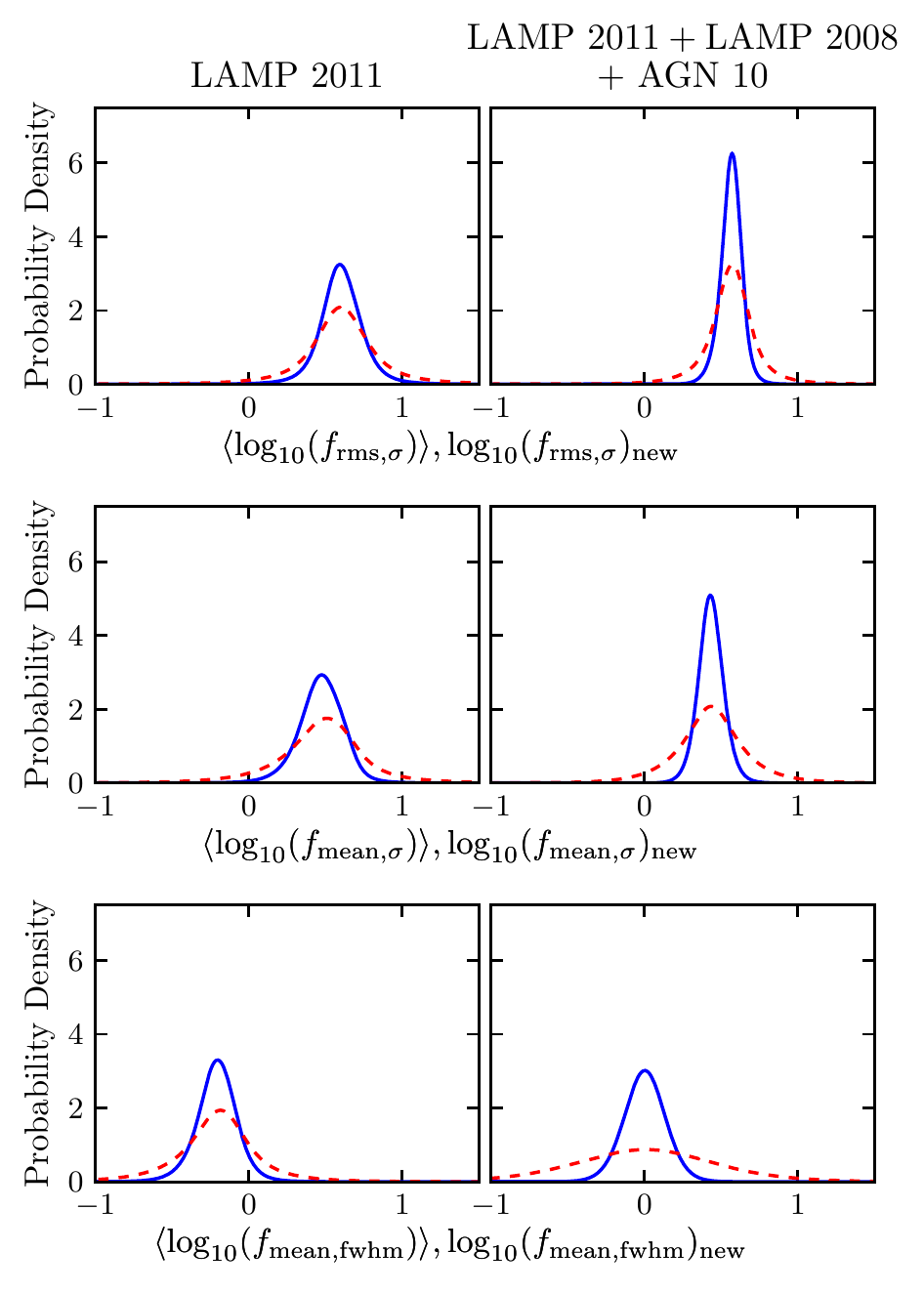}
\caption{The mean scale factor posterior distribution (blue, solid) and the posterior predictive distribution for new scale-factor measurements (red, dashed).
The left-hand panels show the results for the AGNs discussed in this paper.
The right panels show the results when the sample from this paper is combined with the samples from \citetalias{pancoast14b} and \citetalias{Grier++17}.
\label{fig:fmean}}
\end{center}
\end{figure}

To propagate uncertainty, we first assume Gaussian errors in the line width and $\tau$ measurements, with standard deviations given in the respective papers.
We take random draws from these distributions to generate a sample of virial product values of the same size as the black hole mass posterior sample.
The scale factor posterior distribution is then calculated by dividing the black hole mass by the virial product distributions.

\subsubsection{Mean $f$ for the LAMP 2011 Sample}
\label{sect:meanf}

\newcommand{\meanftablecomment}{Summary of the scale factors calculated for the LAMP 2011 sample (L11) and combined LAMP 2011 + \citetalias{pancoast14b} + \citetalias{Grier++17} samples (Comb.).
The mean scale factor is $\log_{10}(\bar{f})$, the dispersion is $\sigma_{\log_{10} f}$, and the mean and standard deviation of the posterior predictive distribution is $\log_{10}(f)_{\rm pred}.$
}
\begin{deluxetable}{c|l|ccc}
\tablecaption{Summary of scale factors}
\tablewidth{0pt}
\tablehead{
\colhead{ } &
\colhead{ } &
\colhead{mean, FWHM} &
\colhead{mean, $\sigma$} &
\colhead{rms, $\sigma$}
}
\startdata
 & $\log_{10}(\bar{f})$ & \lampmeanfmeanfwhm & \lampmeanfmeansigma & \lampmeanfrmssigma \\
L11 & $\sigma_{\log_{10}f}$ & \lampsigmafmeanfwhm & \lampsigmafmeansigma & \lampsigmafrmssigma \\
 & $\log_{10}(f)_{\rm pred}$ & \lamppredfmeanfwhm & \lamppredfmeansigma & \lamppredfrmssigma \\\hline
 & $\log_{10}(\bar{f})$ & \combinedmeanfmeanfwhm & \combinedmeanfmeansigma & \combinedmeanfrmssigma \\
Comb. & $\sigma_{\log_{10}f}$ & \combinedsigmafmeanfwhm & \combinedsigmafmeansigma & \combinedsigmafrmssigma \\
 & $\log_{10}(f)_{\rm pred}$ & \combinedpredfmeanfwhm & \combinedpredfmeansigma & \combinedpredfrmssigma
\enddata
\tablecomments{\meanftablecomment}
\label{tab:table_meanf}
\end{deluxetable}

Here, we examine the mean scale factor for the LAMP 2011 sample as well as the mean scale factor for the sample of all AGNs analyzed using the direct modeling method in this paper.
We model the distribution of scale factors as Gaussian with mean $\log_{10}\bar{f}$ and standard deviation $\sigma_{\log_{10} f}$ given by the dispersion in the individual values of $\log_{10} f$.
The likelihood for different pairs of ($\log_{10}\bar{f}$, $\sigma_{\log_{10} f}$) is calculated using the full posterior PDFs for each AGN.
From this analysis, we can determine both $\log_{10}\bar{f}$ and $\sigma_{\log_{10} f}$ as well as uncertainties in each value.
We also calculate the posterior predictive distribution for our sample which marginalizes over the uncertainty in $\log_{10}\bar{f}$ and $\sigma_{\log_{10} f}$:
\begin{align}
 p(f_{\rm pred} | \mathbf{X}) = \int_{\theta} p(f_{\rm pred} | \theta,\mathbf{X}) p(\theta | \mathbf{X}) d\theta,
\end{align}
where $\mathbf{X}$ is the sample of measurements and $\theta = (\log_{10}\bar{f},\sigma_{\log_{10} f})$.
This is the distribution from which future scale-factor measurements are drawn and is wider than the distribution of our measured scale factors owing to the uncertainty on $\log_{10}\bar{f}$ and $\sigma_{\log_{10} f}$.

The same scale-factor analysis was performed by \citetalias{pancoast14b} (5 objects) and \citetalias{Grier++17} (4 objects).
Here, we combine our sample of 7 objects with those results to expand the sample to 16 objects.
The results for the LAMP 2011 sample and the combined sample of 16 objects are summarized in Table \ref{tab:table_meanf}.
We find that our results for \meanlogfrmssigma\ both in the LAMP 2011 sample and the combined sample are consistent with the results of previous studies \citep[e.g.,][]{graham11,onken04,woo10,woo13,grier13b} to within the quoted uncertainties.
However, we note that \iisa, which has the smallest value for both \logfmeansigma\ and \logfmeanfwhm, does not have a \logfrmssigma\ measurement, owing to the low signal-to-noise ratio in the rms spectrum.
If \logfrmssigma\ for this object has a similarly small value, this could reduce our mean value to be inconsistent with some of the higher measurements.

\begin{figure*}[h!]
\begin{center}
\includegraphics[width=7.0in]{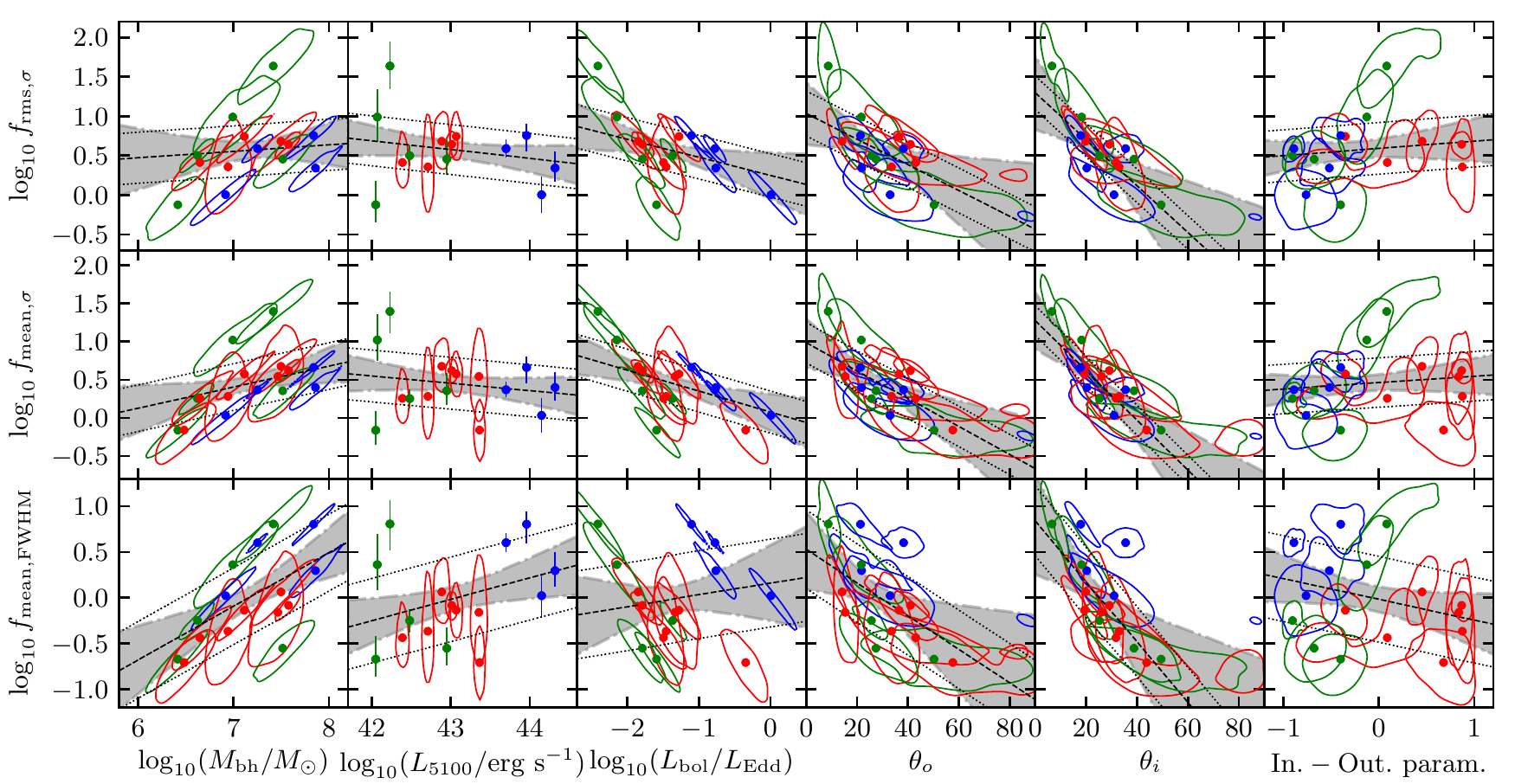}
\caption{Correlations between the scale factor $f$ and select AGNs and model parameters.
The colored dots and contours show the median and 68\% confidence regions of the 2D posterior PDFs for each AGN.
When the abscissa uncertainty is unavailable, the 1D 68\% confidence interval is shown.
The dashed black lines and grey shaded regions give the median and 68\% confidence intervals of the linear regression.
Dotted lines are offset above and below the dashed line by the median value of the intrinsic scatter.
Red points are for the AGNs in this paper, green points are from \citetalias{pancoast14b}, and blue points are from \citetalias{Grier++17}.
\label{fig:fcorrelations}}
\end{center}
\end{figure*}

The posterior predictive distribution gives the distribution from which new $f$ measurements are drawn and so is the appropriate distribution to use when converting new virial product measurements to black hole masses.
The standard deviation of the posterior predictive distribution for \logfrmssigma\ is 0.19, which is half the intrinsic scatter introduced by the $M_{\rm BH} - \sigma_*$ relation \citep{woo10}.
It is possible that the small uncertainty is due to the narrow range in parameter space spanned by the sample used in this analysis, so future measurements will be necessary to solidify this result.
Further, we find a similar value for \logfmeansigma\ as we do for \logfrmssigma, with an intrinsic scatter that is only 0.26.
This suggests that the line dispersion can be measured in the mean spectrum when the rms spectrum is not readily available and still give consistent black hole mass measurements.
Since \iisa\ and \mrkb\ were not included in the calculation of \logfrmssigma$_{\rm pred}$, we re-do this analysis, excluding those two objects from the calculations of \logfmeanfwhm$_{\rm pred}$ and \logfmeansigma$_{\rm pred}$, and find that the scatters remain the same.
This result is particularly useful in cases like \iisa\ and \mrkb\ where the rms spectrum does not have a sufficiently high signal-to-noise ratio to measure a line width.

\subsubsection{Correlations with Other AGN Properties}

\newcommand{\correlationtablecomment}{Results of a fit to the equation $\log_{10} (f) = \alpha + \beta x + N(0,\sigma_{\rm int})$, where $\sigma_{\rm int}$ is the intrinsic scatter and $x$ is the given parameter.}
\begin{deluxetable*}{c|l|cccccc}
\tablecaption{Linear regression results}
\tablewidth{0pt}
\tablehead{
\colhead{$f$-type} &
\colhead{ } &
\colhead{$\log_{10}(M_{\rm bh}/M_{\odot})$ } &
\colhead{$\log_{10}(L_{5100}/{\rm erg~s}^{-1})$ } &
\colhead{$\log_{10}(L_{\rm bol}/L_{\rm Edd})$ } &
\colhead{$\theta_o$ (${\rm deg.}$)} &
\colhead{$\theta_i$ (${\rm deg.}$)} &
\colhead{${\rm In.-Out.~param.}$ }
}
\startdata
\multirow{ 3}{*}{\begin{tabular}{l}mean, \\ FWHM\end{tabular}}  & $\alpha$ & $-4.20_{-2.14}^{+2.18}$ & $-10.1_{-8.1}^{+8.3}$ & $0.16_{-0.41}^{+0.41}$ & $0.53_{-0.44}^{+0.41}$ & $0.84_{-0.59}^{+0.51}$ & $-0.02_{-0.14}^{+0.14}$ \\
 & $\beta$ & $0.59_{-0.30}^{+0.30}$ & $0.23_{-0.19}^{+0.19}$ & $0.13_{-0.29}^{+0.30}$ & $-0.018_{-0.014}^{+0.015}$ & $-0.033_{-0.021}^{+0.023}$ & $-0.23_{-0.24}^{+0.24}$ \\
 & $\sigma_{\rm int}$ & $0.42_{-0.25}^{+0.34}$ & $0.46_{-0.30}^{+0.40}$ & $0.48_{-0.32}^{+0.41}$ & $0.43_{-0.26}^{+0.35}$ & $0.38_{-0.24}^{+0.32}$ & $0.47_{-0.30}^{+0.40}$\\\hline
 & $\alpha$ & $-1.53_{-1.89}^{+1.75}$ & $4.61_{-6.32}^{+6.71}$ & $0.08_{-0.25}^{+0.24}$ & $0.99_{-0.33}^{+0.31}$ & $1.28_{-0.38}^{+0.37}$ & $0.46_{-0.11}^{+0.10}$ \\
${\rm mean},{\sigma}$ & $\beta$ & $0.28_{-0.24}^{+0.26}$ & $-0.10_{-0.16}^{+0.15}$ & $-0.28_{-0.18}^{+0.17}$ & $-0.018_{-0.010}^{+0.011}$ & $-0.033_{-0.016}^{+0.015}$ & $0.08_{-0.18}^{+0.18}$ \\
 & $\sigma_{\rm int}$ & $0.31_{-0.21}^{+0.28}$ & $0.34_{-0.24}^{+0.33}$ & $0.28_{-0.19}^{+0.26}$ & $0.26_{-0.18}^{+0.25}$ & $0.20_{-0.14}^{+0.20}$ & $0.32_{-0.24}^{+0.31}$\\\hline
 & $\alpha$ & $-0.01_{-2.34}^{+2.13}$ & $5.37_{-6.05}^{+6.53}$ & $0.25_{-0.30}^{+0.29}$ & $1.04_{-0.40}^{+0.39}$ & $1.30_{-0.47}^{+0.46}$ & $0.59_{-0.12}^{+0.12}$ \\
${\rm rms},{\sigma}$ & $\beta$ & $0.08_{-0.29}^{+0.32}$ & $-0.11_{-0.15}^{+0.14}$ & $-0.23_{-0.20}^{+0.20}$ & $-0.016_{-0.013}^{+0.013}$ & $-0.030_{-0.019}^{+0.019}$ & $0.09_{-0.19}^{+0.21}$ \\
 & $\sigma_{\rm int}$ & $0.32_{-0.25}^{+0.34}$ & $0.32_{-0.25}^{+0.35}$ & $0.28_{-0.21}^{+0.29}$ & $0.28_{-0.21}^{+0.29}$ & $0.21_{-0.16}^{+0.23}$ & $0.33_{-0.25}^{+0.35}$
\enddata
\tablecomments{\resultstablecomment}
\label{tab:table_correlations}
\end{deluxetable*}

The analysis described in this paper avoids the uncertainty introduced when using an average value for $f$ by modeling the BLR directly.
Unfortunately, such an analysis requires high-quality datasets from long and intensive observing campaigns, limiting its application to small samples of AGNs.
However, using the scale factors found from the direct modeling approach, we can look for correlations between the scale factor and other AGN and BLR properties.
If any exist, this would enable more precise measurements of the scale factor for individual AGNs based on observables.

In Figure \ref{fig:fcorrelations}, we examine correlations between the scale factor $f$ and various model parameters and other observables.
Our calculation of the scale factor depends on the black hole mass inferred by the model, so the uncertainties in the scale factor are intrinsically tied to the uncertainties in other model parameters and values calculated using those parameters.
Further, owing to degeneracies in the model, the uncertainties in other parameters are also correlated with the uncertainties in $f$.
Correlated measurement uncertainties, if not taken into account, will increase (decrease) the measured correlation between two parameters if the sign of the measurement uncertainty correlation is the same as (opposite of) the intrinsic correlation between the two variables.
We use the {\sc IDL} routine \texttt{linmix\_err} \citep{Kelly07} to perform a Bayesian linear regression that accounts for correlated measurement uncertainties.
We should note that \texttt{linmix\_err} assumes Gaussian uncertainties, so the full 2D posterior PDFs for each data point are not used.
The resulting fits are given in Table \ref{tab:table_correlations}.

To quantify the strength of correlation, we compare the median fit slope to the 1$\sigma$ uncertainty in the slope.
We define the following levels of confidence in the existence of a correlation: (0--2)$\sigma$, no evidence; (2--3)$\sigma$, marginal evidence; (3--5)$\sigma$, evidence; $>5\sigma$, conclusive evidence.

Figure \ref{fig:fcorrelations} suggests a possible correlation between \logfmeanfwhm\ and $\log_{10}(M_{\rm BH}/M_\odot)$.
The linear regression analysis finds the slope of the correlation to be $\beta = 0.59^{+0.30}_{-0.30}$, putting it at the boundary between our definitions of ``no evidence'' and ``marginal evidence.''
The presence of this correlation would suggest that the mass of the black hole may influence the shape or dynamics of the BLR, but data spanning a larger range of black hole masses will help confirm or reject the presence of this correlation.
There is no evidence of this correlation for \logfmeansigma\ or \logfrmssigma.

There is also an apparent correlation between \logfmeansigma\ and \logfrmssigma\ and the Eddington fraction in Figure \ref{fig:fcorrelations}.
One might expect this correlation to exist if the accretion rate has a strong influence on the BLR dynamics.
However, since both values are calculated using $M_{\rm BH}$, the measurement uncertainties are highly correlated.
When this is taken into account, there is no evidence of a correlation between the scale factor and Eddington fraction.
We also find no evidence of a correlation between the scale factor and the AGN continuum luminosity at 5100 \AA, $L_{5100}$, whose uncertainties are independent of the uncertainties in $f$.

We do find marginal evidence for a negative correlation between \logfmeansigma\ and $\theta_i$, which was also found by \citetalias{pancoast14b} and \citetalias{Grier++17}.
This result is unsurprising for models similar to ours and is predicted by \citet{goad12}.
For a given disk-like BLR, characteristic of the geometries that we measure, increasing the inclination angle has the effect of increasing the line-of-sight velocity and equivalently the measured line width $v$.
This increases the virial product, requiring a smaller scale factor to recover the same black hole mass.
While the negative correlation appears to also be strong for \logfrmssigma\ and \logfmeanfwhm\ in Figure \ref{fig:fcorrelations}, our full analysis finds no correlation, with $\beta = -0.030^{+0.019}_{-0.019}$ and $-0.033^{+0.023}_{-0.021}$, respectively.

An additional result of this analysis is a measurement of the intrinsic scatter in the relations between the scale factor and other parameters.
We find that for every parameter, the median intrinsic scatter in $f_{{\rm mean},{\rm FWHM}}$ is much higher than that for $f_{{\rm mean},\sigma}$ and $f_{{\rm rms},\sigma}$, albeit with large error bars.
This suggests that the line dispersion provides virial product measurements that are more tightly related to the true black hole mass.
This result is also supported by the results of Section \ref{sect:meanf}, in which the dispersion in the posterior predictive distribution is roughly half the size in the line-dispersion measurements as in the FWHM measurements.
Thus, we suggest that the line dispersion is a more meaningful measure of the line width and should be used when making $M_{\rm BH}$ measurements.


\section{Summary}
\label{sect:summary}

We have analyzed the data of seven AGNs from the Lick AGN Monitoring Project 2011 to constrain the structure and dynamics of the \Hb-emitting BLR. Our results can be summarized as follows.

\begin{enumerate}
    \item The \Hb-emitting BLR is best described by a thick disk that is closer to face-on than edge-on, with a radial distribution that is between Gaussian and exponential, in agreement with the results for the BLRs in \citetalias{pancoast14b} and \citetalias{Grier++17}.
 The \Hb\ emission comes preferentially from the far side of the BLR, which is consistent with photoionization modeling predictions.
    \item Dynamically, the BLR gas can be on elliptical orbits, inflowing or outflowing motions, or a combination of elliptical orbits and either inflowing or outflowing motions.
 The preference for outflowing gas in many of the AGNs is a result that has not been seen in the BLR of other AGNs that have been modeled in this manner.
    \item We measure black hole masses of $\log_{10}(M_{\rm BH}/M_\odot) = $ \iisalogmbh\ for \iisa, \mrkalogmbh\ for \mrka, \mrkblogmbh\ for \mrkb, \mrkclogmbh\ for \mrkc, \mrkdlogmbh\ for \mrkd, \ngcalogmbh\ for \ngca, and \zwialogmbh\ for \zwia.
 All values are fully consistent with previous measurements, except for that of \iisa.
 However, this discrepancy may be due to additional uncertainties in the single-epoch method of black hole mass estimates.
    \item We measure a mean scale factor for the LAMP 2011 sample of \logfrmssigma\ = \lampmeanfrmssigma.
 Combined with the results from \citetalias{pancoast14b} and \citetalias{Grier++17}, we find \logfrmssigma\ = \combinedmeanfrmssigma.
 The posterior predictive distribution for $f$ shows little scatter and can be used for measuring black hole masses with other reverberation mapping data: \logfrmssigma\ = \combinedpredfrmssigma.
 Further, the agreement between these values and the scale factors found by aligning AGNs with the $M_{\rm BH}-\sigma_*$ relation for quiescent galaxies indicates that the $M_{\rm BH}-\sigma_*$ relation for AGN is consistent with that for quiescent galaxies.
    \item The scale factors we recover when using the line dispersion measured in the mean spectrum are consistent with those found when the rms spectrum is used.
 The scatter in the posterior predictive distribution is of similar magnitude, showing that the mean spectrum is a suitable alternative when the rms spectrum is either unavailable or does not have a sufficient signal-to-noise ratio to measure a line width: \logfmeansigma$_{\rm pred}$ = \combinedpredfmeansigma.
 When the FWHM is used instead of the line dispersion, we find the largest scatter, with \logfmeanfwhm$_{\rm pred}$ = \combinedpredfmeanfwhm.
    \item When the line width is measured as the FWHM in the mean spectrum, we find marginal evidence for a correlation between the scale factor \logfmeanfwhm\ and $\log_{10}(M_{\rm BH}/M_\odot)$.
 There is no significant correlation present when the line dispersion is used instead.
 There is also marginal evidence of a correlation between the scale factor and the inclination angle when the line dispersion is used, measured in the mean spectrum.
 We find no significant correlation between the scale factor and the AGN continuum luminosity or Eddington ratio.
\end{enumerate}

The modeling of these objects from the LAMP 2011 sample has nearly doubled the number of AGNs with dynamical modeling of the BLR.
The increased sample has allowed us to measure predictive values of $f$ with scatter that is now smaller than that of the $M_{\rm BH}-\sigma_*$ relation.
These results can be used with traditional reverberation mapping techniques to obtain more precise black hole mass measurements.
However, the significance of correlations between the scale factor and other AGN properties remains uncertain.
Further analysis of data from other reverberation mapping campaigns covering a more diverse range of AGN properties will help uncover these correlations, which can then be used to make accurate $M_{\rm BH}$ measurements on an individual AGN basis.

\acknowledgments
We thank the Lick Observatory staff for their exceptional support during our observing campaign. The Kast spectrograph was made possible through a generous gift from William and Marina Kast. KAIT (at Lick) and its ongoing operation were made possible by donations from Sun Microsystems, Inc., the Hewlett-Packard Company, AutoScope Corporation, Lick Observatory, the National Science Foundation (NSF), the University of California, the Sylvia \& Jim Katzman Foundation, and the TABASGO Foundation. Research at Lick Observatory is partially supported by a generous gift from Google.

The Lick AGN Monitoring Project 2011 is supported by National Science Foundation (NSF) grants AST-1107812, 1107865, 1108665, and 1108835. (Note that findings and conclusions do not necessarily represent views of the NSF.)
Research by P.R.W. and T.T. is supported by NSF grant AST-1412315.
B.J.B. and T.T. acknowledge support from the Packard Foundation through a Packard Fellowship to T.T.
Research by A.J.B. is supported by NSF grant AST-1412693.
The West Mountain Observatory 0.9 m telescope was supported by NSF grant AST–0618209 during this campaign.
M.D.J. would like to thank the Department of Physics and Astronomy at Brigham Young University for continued support of the research efforts at the West Mountain Observatory.
V.N.B. gratefully acknowledges assistance from an NSF Research at Undergraduate Institutions (RUI).
grant AST-1312296.
A.V.F. has been supported by the Christopher R. Redlich Fund, the TABASGO Foundation,  NSF grant AST-1211916, and the Miller Institute for Basic Research in Science (UC Berkeley).
His work was conducted in part at the Aspen Center for Physics, which is 
supported by NSF grant PHY-1607611; he thanks the Center for its hospitality 
during the supermassive black holes workshop in June and July 2018. 
S.F.H. is supported by European Research Council Starting Grant ERC-StG-677117 DUST-IN-THE-WIND.
J.-H.W. acknowledges the support by the National Research Foundation of Korea (NRF) grant funded by the Korea government (No. 2017R1A5A1070354).

\bibliographystyle{apj}
\bibliography{references}

\begin{thebibliography}{}
\expandafter\ifx\csname natexlab\endcsname\relax\def\natexlab#1{#1}\fi

\bibitem[{{Barth} {et~al.}(2011{\natexlab{a}}){Barth}, {Nguyen}, {Malkan},
  {Filippenko}, {Li}, {Gorjian}, {Joner}, {Bennert}, {Botyanszki}, {Cenko},
  {Childress}, {Choi}, {Comerford}, {Cucciara}, {da Silva}, {Duch{\^e}ne},
  {Fumagalli}, {Ganeshalingam}, {Gates}, {Gerke}, {Griffith}, {Harris},
  {Hintz}, {Hsiao}, {Kandrashoff}, {Keel}, {Kirkman}, {Kleiser}, {Laney},
  {Lee}, {Lopez}, {Lowe}, {Moody}, {Morton}, {Nierenberg}, {Nugent},
  {Pancoast}, {Rex}, {Rich}, {Silverman}, {Smith}, {Sonnenfeld}, {Suzuki},
  {Tytler}, {Walsh}, {Woo}, {Yang}, \& {Zeisse}}]{barth11}
{Barth}, A.~J., {Nguyen}, M.~L., {Malkan}, M.~A., {et~al.} 2011{\natexlab{a}},
  \apj, 732, 121

\bibitem[{{Barth} {et~al.}(2011{\natexlab{b}}){Barth}, {Pancoast}, {Thorman},
  {Bennert}, {Sand}, {Li}, {Canalizo}, {Filippenko}, {Gates}, {Greene},
  {Malkan}, {Stern}, {Treu}, {Woo}, {Assef}, {Bae}, {Brewer}, {Buehler},
  {Cenko}, {Clubb}, {Cooper}, {Diamond-Stanic}, {Hiner}, {H{\"o}nig}, {Joner},
  {Kandrashoff}, {Laney}, {Lazarova}, {Nierenberg}, {Park}, {Silverman}, {Son},
  {Sonnenfeld}, {Tollerud}, {Walsh}, {Walters}, {da Silva}, {Fumagalli},
  {Gregg}, {Harris}, {Hsiao}, {Lee}, {Lopez}, {Rex}, {Suzuki}, {Trump},
  {Tytler}, {Worseck}, \& {Yesuf}}]{barth11b}
{Barth}, A.~J., {Pancoast}, A., {Thorman}, S.~J., {et~al.} 2011{\natexlab{b}},
  \apjl, 743, L4

\bibitem[{{Barth} {et~al.}(2015){Barth}, {Bennert}, {Canalizo}, {Filippenko},
  {Gates}, {Greene}, {Li}, {Malkan}, {Pancoast}, {Sand}, {Stern}, {Treu},
  {Woo}, {Assef}, {Bae}, {Brewer}, {Cenko}, {Clubb}, {Cooper},
  {Diamond-Stanic}, {Hiner}, {H{\"o}nig}, {Hsiao}, {Kandrashoff}, {Lazarova},
  {Nierenberg}, {Rex}, {Silverman}, {Tollerud}, \& {Walsh}}]{barth15}
{Barth}, A.~J., {Bennert}, V.~N., {Canalizo}, G., {et~al.} 2015, \apjs, 217, 26

\bibitem[{{Batiste} {et~al.}(2017){Batiste}, {Bentz}, {Raimundo},
  {Vestergaard}, \& {Onken}}]{Batiste++17}
{Batiste}, M., {Bentz}, M.~C., {Raimundo}, S.~I., {Vestergaard}, M., \&
  {Onken}, C.~A. 2017, \apjl, 838, L10

\bibitem[{{Bentz} {et~al.}(2009{\natexlab{a}}){Bentz}, {Peterson}, {Netzer},
  {Pogge}, \& {Vestergaard}}]{bentz09a}
{Bentz}, M.~C., {Peterson}, B.~M., {Netzer}, H., {Pogge}, R.~W., \&
  {Vestergaard}, M. 2009{\natexlab{a}}, \apj, 697, 160

\bibitem[{{Bentz} {et~al.}(2009{\natexlab{b}}){Bentz}, {Walsh}, {Barth},
  {Baliber}, {Bennert}, {Canalizo}, {Filippenko}, {Ganeshalingam}, {Gates},
  {Greene}, {Hidas}, {Hiner}, {Lee}, {Li}, {Malkan}, {Minezaki}, {Sakata},
  {Serduke}, {Silverman}, {Steele}, {Stern}, {Street}, {Thornton}, {Treu},
  {Wang}, {Woo}, \& {Yoshii}}]{bentz09}
{Bentz}, M.~C., {Walsh}, J.~L., {Barth}, A.~J., {et~al.} 2009{\natexlab{b}},
  \apj, 705, 199

\bibitem[{{Bentz} {et~al.}(2010){Bentz}, {Horne}, {Barth}, {Bennert},
  {Canalizo}, {Filippenko}, {Gates}, {Malkan}, {Minezaki}, {Treu}, {Woo}, \&
  {Walsh}}]{bentz10b}
{Bentz}, M.~C., {Horne}, K., {Barth}, A.~J., {et~al.} 2010, \apjl, 720, L46

\bibitem[{{Blandford} \& {McKee}(1982)}]{blandford82}
{Blandford}, R.~D., \& {McKee}, C.~F. 1982, \apj, 255, 419

\bibitem[{{Boroson} \& {Green}(1992)}]{boroson92}
{Boroson}, T.~A., \& {Green}, R.~F. 1992, \apjs, 80, 109

\bibitem[{{Brewer} {et~al.}(2011{\natexlab{a}}){Brewer}, {P{\'a}rtay}, \&
  {Cs{\'a}nyi}}]{brewer11}
{Brewer}, B.~J., {P{\'a}rtay}, L.~B., \& {Cs{\'a}nyi}, G. 2011{\natexlab{a}},
  Statistics and Computing, 21, 649, astrophysics Source Code Library

\bibitem[{{Brewer} {et~al.}(2011{\natexlab{b}}){Brewer}, {Treu}, {Pancoast},
  {Barth}, {Bennert}, {Bentz}, {Filippenko}, {Greene}, {Malkan}, \&
  {Woo}}]{brewer11b}
{Brewer}, B.~J., {Treu}, T., {Pancoast}, A., {et~al.} 2011{\natexlab{b}},
  \apjl, 733, L33

\bibitem[{{Brown} {et~al.}(2013){Brown}, {Baliber}, {Bianco}, {Bowman},
  {Burleson}, {Conway}, {Crellin}, {Depagne}, {De Vera}, {Dilday}, {Dragomir},
  {Dubberley}, {Eastman}, {Elphick}, {Falarski}, {Foale}, {Ford}, {Fulton},
  {Garza}, {Gomez}, {Graham}, {Greene}, {Haldeman}, {Hawkins}, {Haworth},
  {Haynes}, {Hidas}, {Hjelstrom}, {Howell}, {Hygelund}, {Lister}, {Lobdill},
  {Martinez}, {Mullins}, {Norbury}, {Parrent}, {Paulson}, {Petry}, {Pickles},
  {Posner}, {Rosing}, {Ross}, {Sand}, {Saunders}, {Shobbrook}, {Shporer},
  {Street}, {Thomas}, {Tsapras}, {Tufts}, {Valenti}, {Vander Horst}, {Walker},
  {White}, \& {Willis}}]{Brown++13}
{Brown}, T.~M., {Baliber}, N., {Bianco}, F.~B., {et~al.} 2013, \pasp, 125, 1031

\bibitem[{{Busch} {et~al.}(2014){Busch}, {Zuther}, {Valencia-S.}, {Moser},
  {Fischer}, {Eckart}, {Scharw{\"a}chter}, {Gadotti}, \&
  {Wisotzki}}]{Busch++14}
{Busch}, G., {Zuther}, J., {Valencia-S.}, M., {et~al.} 2014, \aap, 561, A140

\bibitem[{{Castell{\'o}-Mor} {et~al.}(2017){Castell{\'o}-Mor}, {Kaspi},
  {Netzer}, {Du}, {Hu}, {Ho}, {Bai}, {Bian}, {Yuan}, \&
  {Wang}}]{Castello-mor++17}
{Castell{\'o}-Mor}, N., {Kaspi}, S., {Netzer}, H., {et~al.} 2017, \mnras, 467,
  1209

\bibitem[{{Cenko} {et~al.}(2006){Cenko}, {Fox}, {Moon}, {Harrison}, {Kulkarni},
  {Henning}, {Guzman}, {Bonati}, {Smith}, {Thicksten}, {Doyle}, {Petrie},
  {Gal-Yam}, {Soderberg}, {Anagnostou}, \& {Laity}}]{cenko06}
{Cenko}, S.~B., {Fox}, D.~B., {Moon}, D.-S., {et~al.} 2006, \pasp, 118, 1396

\bibitem[{{Collin} {et~al.}(2006){Collin}, {Kawaguchi}, {Peterson}, \&
  {Vestergaard}}]{collin06}
{Collin}, S., {Kawaguchi}, T., {Peterson}, B.~M., \& {Vestergaard}, M. 2006,
  \aap, 456, 75

\bibitem[{{Denney} {et~al.}(2009{\natexlab{a}}){Denney}, {Peterson},
  {Dietrich}, {Vestergaard}, \& {Bentz}}]{denney09}
{Denney}, K.~D., {Peterson}, B.~M., {Dietrich}, M., {Vestergaard}, M., \&
  {Bentz}, M.~C. 2009{\natexlab{a}}, \apj, 692, 246

\bibitem[{{Denney} {et~al.}(2009{\natexlab{b}}){Denney}, {Peterson}, {Pogge},
  {Adair}, {Atlee}, {Au-Yong}, {Bentz}, {Bird}, {Brokofsky}, {Chisholm},
  {Comins}, {Dietrich}, {Doroshenko}, {Eastman}, {Efimov}, {Ewald}, {Ferbey},
  {Gaskell}, {Hedrick}, {Jackson}, {Klimanov}, {Klimek}, {Kruse},
  {Lad{\'e}route}, {Lamb}, {Leighly}, {Minezaki}, {Nazarov}, {Onken},
  {Petersen}, {Peterson}, {Poindexter}, {Sakata}, {Schlesinger}, {Sergeev},
  {Skolski}, {Stieglitz}, {Tobin}, {Unterborn}, {Vestergaard}, {Watkins},
  {Watson}, \& {Yoshii}}]{Denney++09}
{Denney}, K.~D., {Peterson}, B.~M., {Pogge}, R.~W., {et~al.}
  2009{\natexlab{b}}, \apjl, 704, L80

\bibitem[{{Denney} {et~al.}(2010){Denney}, {Peterson}, {Pogge}, {Adair},
  {Atlee}, {Au-Yong}, {Bentz}, {Bird}, {Brokofsky}, {Chisholm}, {Comins},
  {Dietrich}, {Doroshenko}, {Eastman}, {Efimov}, {Ewald}, {Ferbey}, {Gaskell},
  {Hedrick}, {Jackson}, {Klimanov}, {Klimek}, {Kruse}, {Lad{\'e}route}, {Lamb},
  {Leighly}, {Minezaki}, {Nazarov}, {Onken}, {Petersen}, {Peterson},
  {Poindexter}, {Sakata}, {Schlesinger}, {Sergeev}, {Skolski}, {Stieglitz},
  {Tobin}, {Unterborn}, {Vestergaard}, {Watkins}, {Watson}, \&
  {Yoshii}}]{denney10}
---. 2010, \apj, 721, 715

\bibitem[{{Du} {et~al.}(2016){Du}, {Lu}, {Hu}, {Qiu}, {Li}, {Huang}, {Wang},
  {Bai}, {Bian}, {Yuan}, {Ho}, {Wang}, \& {SEAMBH Collaboration}}]{Du++16}
{Du}, P., {Lu}, K.-X., {Hu}, C., {et~al.} 2016, \apj, 820, 27

\bibitem[{{Ferland} {et~al.}(1992){Ferland}, {Peterson}, {Horne}, {Welsh}, \&
  {Nahar}}]{Ferland++92}
{Ferland}, G.~J., {Peterson}, B.~M., {Horne}, K., {Welsh}, W.~F., \& {Nahar},
  S.~N. 1992, \apj, 387, 95

\bibitem[{{Ferrarese} \& {Ford}(2005)}]{ferrarese05}
{Ferrarese}, L., \& {Ford}, H. 2005, \ssr, 116, 523

\bibitem[{{Ferrarese} \& {Merritt}(2000)}]{ferrarese00}
{Ferrarese}, L., \& {Merritt}, D. 2000, \apjl, 539, L9

\bibitem[{{Filippenko} {et~al.}(2001){Filippenko}, {Li}, {Treffers}, \&
  {Modjaz}}]{filippenko01}
{Filippenko}, A.~V., {Li}, W.~D., {Treffers}, R.~R., \& {Modjaz}, M. 2001, in
  Astronomical Society of the Pacific Conference Series, Vol. 246, IAU Colloq.
  183: Small Telescope Astronomy on Global Scales, ed. B.~{Paczynski}, W.-P.
  {Chen}, \& C.~{Lemme}, 121

\bibitem[{{Gebhardt} {et~al.}(2000){Gebhardt}, {Bender}, {Bower}, {Dressler},
  {Faber}, {Filippenko}, {Green}, {Grillmair}, {Ho}, {Kormendy}, {Lauer},
  {Magorrian}, {Pinkney}, {Richstone}, \& {Tremaine}}]{gebhardt00}
{Gebhardt}, K., {Bender}, R., {Bower}, G., {et~al.} 2000, \apjl, 539, L13

\bibitem[{{Goad} {et~al.}(2012){Goad}, {Korista}, \& {Ruff}}]{goad12}
{Goad}, M.~R., {Korista}, K.~T., \& {Ruff}, A.~J. 2012, \mnras, 426, 3086

\bibitem[{{Graham} {et~al.}(2011){Graham}, {Onken}, {Athanassoula}, \&
  {Combes}}]{graham11}
{Graham}, A.~W., {Onken}, C.~A., {Athanassoula}, E., \& {Combes}, F. 2011,
  \mnras, 412, 2211

\bibitem[{{Grier} {et~al.}(2017){Grier}, {Pancoast}, {Barth}, {Fausnaugh},
  {Brewer}, {Treu}, \& {Peterson}}]{Grier++17}
{Grier}, C.~J., {Pancoast}, A., {Barth}, A.~J., {et~al.} 2017, \apj, 849, 146

\bibitem[{{Grier} {et~al.}(2013{\natexlab{a}}){Grier}, {Martini}, {Watson},
  {Peterson}, {Bentz}, {Dasyra}, {Dietrich}, {Ferrarese}, {Pogge}, \&
  {Zu}}]{grier13b}
{Grier}, C.~J., {Martini}, P., {Watson}, L.~C., {et~al.} 2013{\natexlab{a}},
  \apj, 773, 90

\bibitem[{{Grier} {et~al.}(2013{\natexlab{b}}){Grier}, {Peterson}, {Horne},
  {Bentz}, {Pogge}, {Denney}, {De Rosa}, {Martini}, {Kochanek}, {Zu},
  {Shappee}, {Siverd}, {Beatty}, {Sergeev}, {Kaspi}, {Araya Salvo}, {Bird},
  {Bord}, {Borman}, {Che}, {Chen}, {Cohen}, {Dietrich}, {Doroshenko}, {Efimov},
  {Free}, {Ginsburg}, {Henderson}, {King}, {Mogren}, {Molina}, {Mosquera},
  {Nazarov}, {Okhmat}, {Pejcha}, {Rafter}, {Shields}, {Skowron}, {Szczygiel},
  {Valluri}, \& {van Saders}}]{grier13a}
{Grier}, C.~J., {Peterson}, B.~M., {Horne}, K., {et~al.} 2013{\natexlab{b}},
  \apj, 764, 47

\bibitem[{{Horne}(1994)}]{horne94}
{Horne}, K. 1994, in Astronomical Society of the Pacific Conference Series,
  Vol.~69, Reverberation Mapping of the Broad-Line Region in Active Galactic
  Nuclei, ed. P.~M. {Gondhalekar}, K.~{Horne}, \& B.~M. {Peterson}, 23--25

\bibitem[{{Horne} {et~al.}(1991){Horne}, {Welsh}, \& {Peterson}}]{horne91}
{Horne}, K., {Welsh}, W.~F., \& {Peterson}, B.~M. 1991, \apjl, 367, L5

\bibitem[{{Kelly}(2007)}]{Kelly07}
{Kelly}, B.~C. 2007, \apj, 665, 1489

\bibitem[{{Kelly} {et~al.}(2009){Kelly}, {Bechtold}, \&
  {Siemiginowska}}]{kelly09}
{Kelly}, B.~C., {Bechtold}, J., \& {Siemiginowska}, A. 2009, \apj, 698, 895

\bibitem[{{Kormendy} \& {Richstone}(1995)}]{kormendy95}
{Kormendy}, J., \& {Richstone}, D. 1995, \araa, 33, 581

\bibitem[{{Kova{\v c}evi{\'c}} {et~al.}(2010){Kova{\v c}evi{\'c}},
  {Popovi{\'c}}, \& {Dimitrijevi{\'c}}}]{kovacevic10}
{Kova{\v c}evi{\'c}}, J., {Popovi{\'c}}, L.~{\v C}., \& {Dimitrijevi{\'c}},
  M.~S. 2010, \apjs, 189, 15

\bibitem[{{Koz{\l}owski}(2016)}]{kozlowski16}
{Koz{\l}owski}, S. 2016, \apj, 826, 118

\bibitem[{{Koz{\l}owski} {et~al.}(2010){Koz{\l}owski}, {Kochanek}, {Udalski},
  {Wyrzykowski}, {Soszy{\'n}ski}, {Szyma{\'n}ski}, {Kubiak}, {Pietrzy{\'n}ski},
  {Szewczyk}, {Ulaczyk}, {Poleski}, \& {OGLE Collaboration}}]{kozlowski10}
{Koz{\l}owski}, S., {Kochanek}, C.~S., {Udalski}, A., {et~al.} 2010, \apj, 708,
  927

\bibitem[{{Li} {et~al.}(2013){Li}, {Wang}, {Ho}, {Du}, \& {Bai}}]{li13}
{Li}, Y.-R., {Wang}, J.-M., {Ho}, L.~C., {Du}, P., \& {Bai}, J.-M. 2013, \apj,
  779, 110

\bibitem[{{Li} {et~al.}(2008){Li}, {Wu}, \& {Wang}}]{Li++08}
{Li}, Z.-Y., {Wu}, X.-B., \& {Wang}, R. 2008, \apj, 688, 826

\bibitem[{{MacLeod} {et~al.}(2010){MacLeod}, {Ivezi{\'c}}, {Kochanek},
  {Koz{\l}owski}, {Kelly}, {Bullock}, {Kimball}, {Sesar}, {Westman}, {Brooks},
  {Gibson}, {Becker}, \& {de Vries}}]{macleod10}
{MacLeod}, C.~L., {Ivezi{\'c}}, {\v Z}., {Kochanek}, C.~S., {et~al.} 2010,
  \apj, 721, 1014

\bibitem[{{Magorrian} {et~al.}(1998){Magorrian}, {Tremaine}, {Richstone},
  {Bender}, {Bower}, {Dressler}, {Faber}, {Gebhardt}, {Green}, {Grillmair},
  {Kormendy}, \& {Lauer}}]{Magorrian++98}
{Magorrian}, J., {Tremaine}, S., {Richstone}, D., {et~al.} 1998, \aj, 115, 2285

\bibitem[{{Maoz} {et~al.}(1990){Maoz}, {Netzer}, {Leibowitz}, {Brosch}, {Laor},
  {Mendelson}, {Beck}, {Almoznino}, \& {Mazeh}}]{Mao++90}
{Maoz}, D., {Netzer}, H., {Leibowitz}, E., {et~al.} 1990, \apj, 351, 75

\bibitem[{{Marconi} {et~al.}(2008){Marconi}, {Axon}, {Maiolino}, {Nagao},
  {Pastorini}, {Pietrini}, {Robinson}, \& {Torricelli}}]{marconi08}
{Marconi}, A., {Axon}, D.~J., {Maiolino}, R., {et~al.} 2008, \apj, 678, 693

\bibitem[{{Marconi} {et~al.}(2009){Marconi}, {Axon}, {Maiolino}, {Nagao},
  {Pietrini}, {Risaliti}, {Robinson}, \& {Torricelli}}]{marconi09}
---. 2009, \apjl, 698, L103

\bibitem[{{Markwardt}(2009)}]{Markwardt09}
{Markwardt}, C.~B. 2009, in Astronomical Society of the Pacific Conference
  Series, Vol. 411, Astronomical Data Analysis Software and Systems XVIII, ed.
  D.~A. {Bohlender}, D.~{Durand}, \& P.~{Dowler}, 251

\bibitem[{{Netzer}(2009)}]{netzer09}
{Netzer}, H. 2009, \apj, 695, 793

\bibitem[{{Netzer} \& {Marziani}(2010)}]{netzer10}
{Netzer}, H., \& {Marziani}, P. 2010, \apj, 724, 318

\bibitem[{{O'Brien} {et~al.}(1994){O'Brien}, {Goad}, \&
  {Gondhalekar}}]{OBrien++94}
{O'Brien}, P.~T., {Goad}, M.~R., \& {Gondhalekar}, P.~M. 1994, \mnras, 268, 845

\bibitem[{{Onken} {et~al.}(2004){Onken}, {Ferrarese}, {Merritt}, {Peterson},
  {Pogge}, {Vestergaard}, \& {Wandel}}]{onken04}
{Onken}, C.~A., {Ferrarese}, L., {Merritt}, D., {et~al.} 2004, \apj, 615, 645

\bibitem[{{Pancoast} {et~al.}(2011){Pancoast}, {Brewer}, \&
  {Treu}}]{pancoast11}
{Pancoast}, A., {Brewer}, B.~J., \& {Treu}, T. 2011, \apj, 730, 139

\bibitem[{{Pancoast} {et~al.}(2014{\natexlab{a}}){Pancoast}, {Brewer}, \&
  {Treu}}]{pancoast14a}
---. 2014{\natexlab{a}}, \mnras, 445, 3055

\bibitem[{{Pancoast} {et~al.}(2014{\natexlab{b}}){Pancoast}, {Brewer}, {Treu},
  {Park}, {Barth}, {Bentz}, \& {Woo}}]{pancoast14b}
{Pancoast}, A., {Brewer}, B.~J., {Treu}, T., {et~al.} 2014{\natexlab{b}},
  \mnras, 445, 3073

\bibitem[{{Pancoast} {et~al.}(2012){Pancoast}, {Brewer}, {Treu}, {Barth},
  {Bennert}, {Canalizo}, {Filippenko}, {Gates}, {Greene}, {Li}, {Malkan},
  {Sand}, {Stern}, {Woo}, {Assef}, {Bae}, {Buehler}, {Cenko}, {Clubb},
  {Cooper}, {Diamond-Stanic}, {Hiner}, {H{\"o}nig}, {Joner}, {Kandrashoff},
  {Laney}, {Lazarova}, {Nierenberg}, {Park}, {Silverman}, {Son}, {Sonnenfeld},
  {Thorman}, {Tollerud}, {Walsh}, \& {Walters}}]{pancoast12}
---. 2012, \apj, 754, 49

\bibitem[{{Park} {et~al.}(2012){Park}, {Kelly}, {Woo}, \& {Treu}}]{park12b}
{Park}, D., {Kelly}, B.~C., {Woo}, J.-H., \& {Treu}, T. 2012, \apjs, 203, 6

\bibitem[{{Pei} {et~al.}(2017){Pei}, {Fausnaugh}, {Barth}, {Peterson}, {Bentz},
  {De Rosa}, {Denney}, {Goad}, {Kochanek}, {Korista}, {Kriss}, {Pogge},
  {Bennert}, {Brotherton}, {Clubb}, {Dalla Bont{\`a}}, {Filippenko}, {Greene},
  {Grier}, {Vestergaard}, {Zheng}, {Adams}, {Beatty}, {Bigley}, {Brown},
  {Brown}, {Canalizo}, {Comerford}, {Coker}, {Corsini}, {Croft}, {Croxall},
  {Deason}, {Eracleous}, {Fox}, {Gates}, {Henderson}, {Holmbeck}, {Holoien},
  {Jensen}, {Johnson}, {Kelly}, {Kim}, {King}, {Lau}, {Li}, {Lochhaas}, {Ma},
  {Manne-Nicholas}, {Mauerhan}, {Malkan}, {McGurk}, {Morelli}, {Mosquera},
  {Mudd}, {Muller Sanchez}, {Nguyen}, {Ochner}, {Ou-Yang}, {Pancoast}, {Penny},
  {Pizzella}, {Poleski}, {Runnoe}, {Scott}, {Schimoia}, {Shappee}, {Shivvers},
  {Simonian}, {Siviero}, {Somers}, {Stevens}, {Strauss}, {Tayar}, {Tejos},
  {Treu}, {Van Saders}, {Vican}, {Villanueva}, {Yuk}, {Zakamska}, {Zhu},
  {Anderson}, {Ar{\'e}valo}, {Bazhaw}, {Bisogni}, {Borman}, {Bottorff},
  {Brandt}, {Breeveld}, {Cackett}, {Carini}, {Crenshaw}, {De
  Lorenzo-C{\'a}ceres}, {Dietrich}, {Edelson}, {Efimova}, {Ely}, {Evans},
  {Ferland}, {Flatland}, {Gehrels}, {Geier}, {Gelbord}, {Grupe}, {Gupta},
  {Hall}, {Hicks}, {Horenstein}, {Horne}, {Hutchison}, {Im}, {Joner}, {Jones},
  {Kaastra}, {Kaspi}, {Kelly}, {Kennea}, {Kim}, {Kim}, {Klimanov}, {Lee},
  {Leonard}, {Lira}, {MacInnis}, {Mathur}, {McHardy}, {Montouri}, {Musso},
  {Nazarov}, {Netzer}, {Norris}, {Nousek}, {Okhmat}, {Papadakis}, {Parks},
  {Pott}, {Rafter}, {Rix}, {Saylor}, {Schn{\"u}lle}, {Sergeev}, {Siegel},
  {Skielboe}, {Spencer}, {Starkey}, {Sung}, {Teems}, {Turner}, {Uttley},
  {Villforth}, {Weiss}, {Woo}, {Yan}, {Young}, \& {Zu}}]{Pei++17}
{Pei}, L., {Fausnaugh}, M.~M., {Barth}, A.~J., {et~al.} 2017, \apj, 837, 131

\bibitem[{{Peterson}(1993)}]{peterson93}
{Peterson}, B.~M. 1993, \pasp, 105, 247

\bibitem[{{Peterson} {et~al.}(2004){Peterson}, {Ferrarese}, {Gilbert}, {Kaspi},
  {Malkan}, {Maoz}, {Merritt}, {Netzer}, {Onken}, {Pogge}, {Vestergaard}, \&
  {Wandel}}]{Pet++04}
{Peterson}, B.~M., {Ferrarese}, L., {Gilbert}, K.~M., {et~al.} 2004, \apj, 613,
  682

\bibitem[{{Santos-Lle{\'o}} {et~al.}(2001){Santos-Lle{\'o}}, {Clavel},
  {Schulz}, {Altieri}, {Barr}, {Alloin}, {Berlind}, {Bertram}, {Crenshaw},
  {Edelson}, {Giveon}, {Horne}, {Huchra}, {Kaspi}, {Kriss}, {Krolik}, {Malkan},
  {Malkov}, {Netzer}, {O'Brien}, {Peterson}, {Pogge}, {Pronik}, {Qian},
  {Reichert}, {Rodr{\'{\i}}guez-Pascual}, {Sergeev}, {Tao}, {Tokarz}, {Wagner},
  {Wamsteker}, \& {Wilkes}}]{San++01}
{Santos-Lle{\'o}}, M., {Clavel}, J., {Schulz}, B., {et~al.} 2001, \aap, 369, 57

\bibitem[{{Schulze} \& {Wisotzki}(2010)}]{Schulze+10}
{Schulze}, A., \& {Wisotzki}, L. 2010, \aap, 516, A87

\bibitem[{{Shen} \& {Kelly}(2010)}]{Shen+10}
{Shen}, Y., \& {Kelly}, B.~C. 2010, \apj, 713, 41

\bibitem[{{Skielboe} {et~al.}(2015){Skielboe}, {Pancoast}, {Treu}, {Park},
  {Barth}, \& {Bentz}}]{skielboe15}
{Skielboe}, A., {Pancoast}, A., {Treu}, T., {et~al.} 2015, \mnras, 454, 144

\bibitem[{{van Groningen} \& {Wanders}(1992)}]{vanGron+92}
{van Groningen}, E., \& {Wanders}, I. 1992, \pasp, 104, 700

\bibitem[{{V{\'e}ron-Cetty} {et~al.}(2004){V{\'e}ron-Cetty}, {Joly}, \&
  {V{\'e}ron}}]{veron-cetty04}
{V{\'e}ron-Cetty}, M.-P., {Joly}, M., \& {V{\'e}ron}, P. 2004, \aap, 417, 515

\bibitem[{{Walsh} {et~al.}(2009){Walsh}, {Minezaki}, {Bentz}, {Barth},
  {Baliber}, {Li}, {Stern}, {Bennert}, {Brown}, {Canalizo}, {Filippenko},
  {Gates}, {Greene}, {Malkan}, {Sakata}, {Street}, {Treu}, {Woo}, \&
  {Yoshii}}]{walsh09}
{Walsh}, J.~L., {Minezaki}, T., {Bentz}, M.~C., {et~al.} 2009, \apjs, 185, 156

\bibitem[{{Welsh} \& {Horne}(1991)}]{welsh91}
{Welsh}, W.~F., \& {Horne}, K. 1991, \apj, 379, 586

\bibitem[{{Whittle}(1992)}]{whittle92}
{Whittle}, M. 1992, \apjs, 79, 49

\bibitem[{{Woo} {et~al.}(2013){Woo}, {Schulze}, {Park}, {Kang}, {Kim}, \&
  {Riechers}}]{woo13}
{Woo}, J.-H., {Schulze}, A., {Park}, D., {et~al.} 2013, \apj, 772, 49

\bibitem[{{Woo} {et~al.}(2010){Woo}, {Treu}, {Barth}, {Wright}, {Walsh},
  {Bentz}, {Martini}, {Bennert}, {Canalizo}, {Filippenko}, {Gates}, {Greene},
  {Li}, {Malkan}, {Stern}, \& {Minezaki}}]{woo10}
{Woo}, J.-H., {Treu}, T., {Barth}, A.~J., {et~al.} 2010, \apj, 716, 269

\bibitem[{{Zu} {et~al.}(2013){Zu}, {Kochanek}, {Koz{\l}owski}, \&
  {Udalski}}]{zu13}
{Zu}, Y., {Kochanek}, C.~S., {Koz{\l}owski}, S., \& {Udalski}, A. 2013, \apj,
  765, 106

\bibitem[{{Zu} {et~al.}(2011){Zu}, {Kochanek}, \& {Peterson}}]{zu11}
{Zu}, Y., {Kochanek}, C.~S., \& {Peterson}, B.~M. 2011, \apj, 735, 80

\end{thebibliography}

\end{document}